\documentclass[aps,superscriptaddress,prc,twocolumn,nofootinbib,showpacs,showkeys]{revtex4-1}
\usepackage{graphicx}
\usepackage{epstopdf}
\usepackage{subfigure}
\usepackage{epsfig}
\usepackage{amsmath,amssymb,amsfonts}
\usepackage[utf8]{inputenc}
\usepackage{color}
\usepackage[normalem]{ulem}
\usepackage[bookmarks,
            bookmarksopen = true,
            bookmarksnumbered = true,
            linktocpage,
            colorlinks = true,
            linkcolor = blue,
            urlcolor  = blue,
            citecolor = blue,
            anchorcolor = green,
            hyperindex = true,
            hyperfigures]
            {hyperref}
\allowdisplaybreaks
\graphicspath{{./Figures/}}

\begin{document}

\title{Suppression, flow, and chemical components of heavy flavor hadrons in relativistic heavy-ion collisions within an improved Langevin framework}

\author{Shu-Qing Li}
\affiliation{School of Physics and Electronic Engineering, Jining University, Qufu, Shandong, 273155, China}
\affiliation{School of Physics, Southeast University, Nanjing, Jiangsu, 211189, China}
\author{Wen-Jing Xing}
\email{wenjing.xing@usc.edu.cn}
\affiliation{School of Nuclear Science and Technology, University of South China, Hengyang, Hunan, 421001, China}
\author{Shanshan Cao}
\email{shanshan.cao@sdu.edu.cn}
\affiliation{School of Physics, Southeast University, Nanjing, Jiangsu, 211189, China}
\author{Guang-You Qin}
\email{guangyou.qin@mail.ccnu.edu.cn}
\affiliation{Institute of Particle Physics and Key Laboratory of Quark and Lepton Physics (MOE), Central China Normal University, Wuhan, Hubei, 430079, China}



\begin{abstract}
We improve the Langevin model to incorporate both perturbative and non-perturbative interactions between heavy quarks and the quark-gluon plasma (QGP) in relativistic heavy-ion collisions. The transport coefficient of heavy quarks is evaluated by a potential scattering model in which both the short-range Yukawa interaction and the long-range string interaction are included. The updated model provides the correct thermal equilibrium limit of the heavy quark system with the medium background, and improves the accuracy in describing the nuclear modification factor and elliptic flow coefficient of $D$ mesons observed at RHIC and the LHC. Nuclear modification of different chemical components of charm hadrons, including $D^0$, $D_s^+$, $\Lambda_c^+$, $\Xi_c^0$, and $\Omega_c^0$, is investigated. Their spectrum ratios are found sensitive to not only the hadronization mechanism of charm quarks but also the medium-modified charm quark spectrum in heavy-ion collisions, offering additional constraints on the heavy quark dynamics inside a color-deconfined nuclear medium.
\end{abstract}

\maketitle


\section{Introduction}
\label{sec:Introduction}

Heavy quarks serve as a clean probe of the quark-gluon plasma (QGP) created in high-energy nuclear collisions. They are predominantly born from the initial hard partonic scatterings and then witness the full evolution history of the QGP medium. Heavy flavor hadrons at different momentum scales probe different aspects of heavy-quark-medium interactions: at low transverse momentum (\(p_\mathrm{T}\)), they reveal the diffusion and thermalization processes of heavy quarks inside the QGP~\cite{Moore:2004tg,vanHees:2004gq,Gossiaux:2006yu,Cao:2011et,Prakash:2026bog}; at intermediate \(p_\mathrm{T}\), they are sensitive to the hadronization process from free quark states to hadron bound states~\cite{Plumari:2017ntm,Song:2018tpv,He:2019vgs,Cho:2019lxb,Cao:2019iqs,Xu:2025ivv}; and at high \(p_\mathrm{T}\), they enable studies on the flavor and mass dependence of jet quenching through a dense medium~\cite{Zhang:2003wk,Djordjevic:2013pba,Huang:2013vaa,Kang:2016ofv,Zhang:2018nie,Du:2018yuf,Xing:2019xae,Dang:2023tmb,Wang:2024yag}. Over the past two decades, significant efforts have been devoted to understanding the nuclear modification of heavy flavor hadrons in heavy-ion collisions. One may find detailed reviews and model comparisons about heavy quarks in Refs.~\cite{Dong:2019byy,Dong:2019unq,Cao:2018ews,Rapp:2018qla,Zhao:2023nrz}.

By incorporating elastic and inelastic scatterings between heavy quarks and the QGP, perturbative calculations provide a successful description of the heavy flavor data at high $p_\mathrm{T}$ (above 10~GeV)~\cite{Gossiaux:2010yx,Uphoff:2012gb,Cao:2016gvr,Ke:2018tsh,Xing:2019xae,Liu:2021izt}. To the contrary, it has been explicitly shown that the perturbative method is insufficient for heavy quark dynamics at low \(p_\mathrm{T}\)~\cite{CaronHuot:2008uh}: the next-to-leading order result on the heavy quark transport coefficient can be different from the leading-order result by a factor of 5 when a reasonable strong coupling strength is applied. To describe the heavy flavor data using perturbative-driven models, one needs to apply a \(K\) factor to enhance the scattering rate of heavy quarks~\cite{Nahrgang:2013saa,Cao:2016gvr} or mimic non-perturbative effects using a temperature and (or) momentum dependent running coupling strength~\cite{Gossiaux:2010yx,Das:2015ana}. A more delicate treatment of non-perturbative scatterings is to employ a Cornell-type potential to describe the interaction between a heavy quark and a thermal medium, as implemented in the \(T\)-matrix approach~\cite{vanHees:2007me,Riek:2010fk,He:2011qa,Liu:2018syc}. This potential contains a short-range Yukawa part for the perturbative interaction and a long-range string part for the non-perturbative interaction, and can be constrained by the lattice QCD data on the color potential at finite temperature. While the \(T\)-matrix approach sums interactions at all order, the Born approximation (leading order) of this potential scattering picture is implemented in the linear Boltzmann transport (LBT) model~\cite{Xing:2021xwc}, which also significantly improves the description of the nuclear modification factor (\(R_\mathrm{AA}\)) and elliptic flow coefficient (\(v_2\)) of $D$ mesons at low to intermediate $p_\mathrm{T}$. Hadronization is another typical non-perturbative process that strongly affects the heavy flavor observables. While high momentum heavy quarks mainly form hadrons via fragmentation, which is well constrained in electron-positron and proton-proton collisions, low momentum heavy quarks can combine with thermal partons inside the QGP into hadrons, which is called coalescence or recombination. Although large uncertainties still exist in coalescence models~\cite{Zhao:2023nrz}, it is commonly agreed that the coalescence process is essential for understanding the baryon-to-meson ratio ($\Lambda_c^+/D^0$) and strange-to-non-strange ratio ($D_s^+/D^0$) in heavy flavor hadrons at low to intermediate $p_\mathrm{T}$ observed at RHIC~\cite{STAR:2019ank,STAR:2021tte} and the LHC~\cite{Acharya:2018hre,Acharya:2018ckj,ALICE:2022wpn}. Recently, the ALICE Collaboration released the first measurement on the \(\Xi_c^0\) production in heavy-ion collisions, further challenging the precision of heavy quark models~\cite{ALICE:2026fzw}. So far, no existing model quantitatively describes the \(\Xi_c^0/D^0\), \(\Xi_c^0/\Lambda_c^+\), and \(\Xi_c^0/D_s^+\) ratios measured by ALICE. 

In this work, we improve the modified Langevin model~\cite{Cao:2013ita,Cao:2015hia,Li:2020kax} for heavy quark interaction with the QGP. This model was developed in our earlier studies to simultaneously incorporate elastic and inelastic scatterings of heavy quarks inside the QGP, but a constant $D_\mathrm{s}(2\pi T)$ was used as a model parameter, with $D_\mathrm{s}$ the spatial diffusion coefficient of heavy quarks and $T$ the medium temperature. Here, we introduce the energy and temperature dependence of this parameter, evaluated via the potential scattering model with both perturbative and non-perturbative interactions included~\cite{Xing:2021xwc}. With the non-perturbative effect taken into account, significant improvement is achieved in describing the heavy flavor data at RHIC and the LHC. This Langevin model provides consistent results with the LBT model implementing the same potential scattering method~\cite{Xing:2021xwc}, while being much more computationally efficient compared to the LBT model regarding heavy flavor observables. Therefore, it is an ideal tool for extracting the heavy quark transport coefficient or the color potential between quark pairs from the heavy flavor data in the future. In addition, we extend the population of heavy flavor hadrons in our calculations, from $D^0$, $D_s^+$, $\Lambda_c^+$ to $\Xi_c^0$ and $\Omega_c^0$, and show that the charm hadron chemistry (spectrum ratio between different charm hadrons) not only depends on the hadronization mechanism, but also relies on the medium-modified charm quark spectrum in heavy-ion collisions, offering further constraints on the heavy quark dynamics inside the QGP. 

The remainder of this paper is organized as follows. In Sec.~\ref{sec:Theoretical framework}, we develop the Langevin model to incorporate a heavy quark transport coefficient evaluated by the potential scattering model, and examine the thermal equilibrium limit approached by the heavy quark system. In Sec.~\ref{sec:results}, by coupling this improved Langevin model with a realistic initial condition of heavy quarks and a hydrodynamic simulation of the QGP medium, we calculate the spectrum suppression, elliptic flow, and chemistry of charm hadrons, and investigate their dependence on the interaction dynamics between heavy quarks and the QGP. In the end, we summarize in Sec.~\ref{sec:summary}.


\section{Theoretical framework}
\label{sec:Theoretical framework}

\subsection{Modified Langevin equation}
\label{sec:LangevinEquation}

We use a modified Langevin equation~\cite{Cao:2013ita} to describe the time ($t$) evolution of the heavy quark momentum  ($\vec{p}$) inside a QGP medium, which extends the conventional Langevin equation for multiple quasi-elastic scatterings between a heavy quark and the QGP medium to include medium-induced gluon emissions from the heavy quark:
\begin{equation}
\label{eq:Langevin}
\frac{d\vec{p}}{dt} = -\eta_\mathrm{D}(p)\,\vec{p} + \vec{\xi} + \vec{f}_{g}.
\end{equation}
The first two terms on the right-hand side are inherited from the conventional Langevin equation, representing the drag force and the thermal random force experienced by the heavy quark due to its multiple soft quasi-elastic scatterings with the medium background. The third term is introduced to describe the recoil force exerted on the heavy quark as it emits gluons. We assume the thermal random force follows the correlation function of white noise as 
\begin{equation}
\langle \xi^{i}(t)\xi^{j}(t^{\prime}) \rangle = \kappa\,\delta^{ij}\delta(t-t^{\prime}),
\end{equation}
where \(\kappa\) is the momentum space diffusion coefficient of heavy quarks. 

In our earlier work, $\kappa$ was assumed to be independent of the heavy quark momentum, and therefore was related to the spatial diffusion coefficient ($D_\mathrm{s}$) and the drag coefficient ($\eta_\mathrm{D}$) of heavy quarks via
\begin{equation}
\label{eq:DsSimplify}
D_{\mathrm{s}} = \frac{T}{E\eta_\mathrm{D}} = \frac{2T^{2}}{\kappa},
\end{equation}
in which $E$ represents the heavy quark energy, $T$ represents the medium temperature, and the fluctuation-dissipation relation \(\eta_\mathrm{D}=\kappa/(2TE)\) is used. A temperature and momentum independent $D_\mathrm{s} (2\pi T)$ was treated as a model parameter fitted to various heavy flavor data~\cite{Cao:2013ita,Cao:2015hia,Li:2020kax,Li:2021xbd,Li:2024wqq}. In this work, we evaluate the $\kappa$ coefficient of heavy quarks scattering with a Cornell-type potential, in which both perturbative and non-perturbative effects are taken into account. Due to the momentum (or energy) dependence of $\kappa$, the fluctuation dissipation theorem needs to be generalized to 
\begin{equation}
\label{eq:generalizedFDT}
2\eta_\mathrm{D}ET - \kappa + (1-\xi)T\frac{\partial \kappa}{\partial E} = 0,
\end{equation}
and the spatial diffusion coefficient is related to the drag coefficient via
\begin{equation}
\label{eq:SpatialDiffusionCoefficient}
D_{\mathrm{s}} = \frac{\kappa}{2E^{2}\eta_\mathrm{D}^{2}}.
\end{equation}
Here, the parameter $\xi$ characterizes how the Langevin equation, Eq.~(\ref{eq:Langevin}), is discretized when being solved numerically:
\begin{equation}
\label{eq:discretizedLangevin}
\begin{aligned}
\vec{p}(t+\Delta t) &= \vec{p}(t)-\eta_\mathrm{D}[\vec{p}(t)+\xi\Delta\vec{p})]\,\vec{p}(t) \Delta t \\
&+ \vec{\xi}[\vec{p}(t)+\xi\Delta\vec{p}]  \Delta t + \vec{f}_{g}[\vec{p}(t)+\xi\Delta\vec{p}] \Delta t,
\end{aligned}
\end{equation}
with $\Delta \vec{p}$ given by $\vec{p}(t+\Delta t)-\vec{p}(t)$. We adopt the pre-point Ito scheme, corresponding to $\xi=0$. Therefore, with the knowledge of heavy quark momentum $\vec{p}$ for a given time step, one may evaluate $\kappa$, $\partial\kappa/\partial E$, and thus $\eta_\mathrm{D}$ via Eq.~(\ref{eq:generalizedFDT}) for this time step, and further obtain $\vec{p}$ for the next time step. The spatial diffusion coefficient \(D_{\mathrm{s}}(2\pi T)\) is no longer a constant parameter in the present study, but rather a derived quantity that depends on both the medium temperature and the heavy quark momentum. One may verify that if $\kappa$ is energy independent, Eqs.~(\ref{eq:generalizedFDT}) and~(\ref{eq:SpatialDiffusionCoefficient}) reduce to Eq.~(\ref{eq:DsSimplify}). 

The third term in Eq.~(\ref{eq:Langevin}) is related to the momentum of emitted gluons per unit time as \(\vec{f}_{g}=-d\vec{p}_{g}/dt\). With a sufficiently small $\Delta t$, the gluon emission probability within \((t,t+\Delta t)\) can be given by the average number of emitted gluons within this time step as
\begin{equation}
\label{eq:gluonProb}
P_{\mathrm{rad}}(t,\Delta t) = \langle N_{g}(t,\Delta t)\rangle
= \Delta t \int dx dk_{\perp}^{2}
\frac{dN_{g}}{dx dk_{\perp}^{2} dt}.
\end{equation}
In our model, the medium-induced gluon spectrum is taken from the higher-twist energy loss calculation as~\cite{Guo:2000nz,Majumder:2009ge,Zhang:2003wk}
\begin{equation}
\label{eq:gluonSpectrum}
\frac{dN_{g}}{dx dk_{\perp}^{2} dt}=
\frac{2\alpha_\mathrm{s}C_A P(x)\hat{q}}{\pi k_{\perp}^{4}}
\left(\frac{k_{\perp}^{2}}{k_{\perp}^{2}+x^{2}m_Q^{2}}\right)^{4}
\sin^{2}\!\left(\frac{t-t_{i}}{2\tau_{f}}\right),
\end{equation}
where \(x\) is the fractional energy taken by an emitted gluon from its parent heavy quark, \(k_{\perp}\) is its transverse momentum relative to the heavy quark, \(\alpha_\mathrm{s}\) is the strong coupling coefficient running with \(k_{\perp}\) as $\alpha_\mathrm{s} = 4\pi/[9\ln(k_{\perp}^2/\Lambda_{\rm QCD}^2)]$ with $\Lambda_{\rm QCD}=0.2$~GeV,  $C_A=3$ is the color factor, \(P(x)\) is the \(Q\rightarrow Qg\) splitting function, $m_Q$ is the heavy quark mass, $t_i$ is the production time of the heavy quark or the time of its previous splitting, and
$\tau_{f} = 2Ex(1-x)/(k_{\perp}^{2}+x^{2}m_Q^{2})$ is the gluon formation time. The quartic term in the above equation is known as the ``dead cone factor'', characterizing the mass dependence of the radiative energy loss of a hard parton. In addition, \(\hat{q}\) is the jet transport coefficient, quantifying the transverse momentum broadening of the hard parton caused by elastic scatterings. In this work, it is related to the momentum space diffusion of heavy quarks by a dimension factor as $\hat{q} = 2\kappa$. 
The notations presented here are consistent with those used in the LBT model~\cite{Cao:2016gvr,Xing:2021xwc}: the color factor of the hard parton is taken out of the hard parton's splitting function \(P(x)\) and absorbed into its transport coefficient 
$\hat{q}$. This convention appears different from, though consistent with, that used in Refs.~\cite{Cao:2013ita,Cao:2015hia}. For a given time step, we first use Eq.~(\ref{eq:gluonProb}) to determine whether gluon emission occurs. If it occurs, we use Eq.~(\ref{eq:gluonSpectrum}) to sample gluon momentum $\vec{p}_g$ and use  $-\vec{p}_g$ as the last term of Eq.~(\ref{eq:discretizedLangevin}) for this time step. Since the gluon absorption process is not included in our model, we set a lower cutoff of the emitted gluon energy---$x_\mathrm{min}=\pi T/E$---to mimic the detailed balance between gluon emission and absorption below the thermal scale. Later, we will show that heavy quarks can approach the correct thermal equilibrium limit with the medium within our model.

\subsection{Transport coefficients of heavy quarks}
\label{sec:Langevin_PNP}

A key improvement of our Langevin model in this work is that instead of parameterizing the spatial diffusion coefficient ($D_\mathrm{s}$) of heavy quarks, we evaluate their momentum space diffusion coefficient ($\kappa$) by considering both perturbative and non-perturbative scatterings between heavy quarks and the QGP medium. Here, we briefly summarize our methodology developed in Ref.~\cite{Xing:2021xwc} for evaluating this $\kappa$.

We consider a heavy quark scattering with a thermal quark inside the QGP medium via a Cornell-type potential~\cite{Liu:2017qah} parametrized as
\begin{equation}
\label{eq:CornellPotential}
V(r)=V_{\mathrm{Y}}(r)+V_{\mathrm{S}}(r)=
-\frac{4}{3}\alpha_\mathrm{s}\frac{e^{-m_{d}r}}{r}
-\frac{\sigma e^{-m_{s}r}}{m_{s}},
\end{equation}
with $r$ the distance between the two quarks. The Yukawa potential in the first term represents the short-range perturbative interaction, and the string potential in the second term represents the long-range non-perturbative interaction. The parameters in Eq.~(\ref{eq:CornellPotential}) take the values of \(\alpha_\mathrm{s}=0.27\), \(\sigma=0.45~\mathrm{GeV}^{2}\), \(m_{d}=2T+0.2~\mathrm{GeV}\), and \(m_{s}=\sqrt{0.1~\mathrm{GeV}\times T}\), which were fitted to the nuclear modification data of $D$ mesons at RHIC and the LHC within the LBT model and shown to reproduce the $V(r)$ predicted by the lattice QCD calculation~\cite{Xing:2021xwc}.

Via a Fourier transformation, one obtains the potential in the momentum space as
\begin{equation}
\label{eq:CornellPotentialMomentumSpace}
\begin{aligned}
V(\vec{q})&=V_{\mathrm{Y}}(\vec{q})+V_{\mathrm{S}}(\vec{q})\\
&=-\frac{4\pi\alpha_\mathrm{s}C_{F}}{m_{d}^{2}+|\vec{q}|^{2}}
-\frac{8\pi\sigma}{(m_{s}^{2}+|\vec{q}|^{2})^{2}},
\end{aligned}
\end{equation}
in which $C_F=4/3$ is the color factor and \(\vec{q}\) is the momentum transfer between the two quarks. This $V(\vec{q})$ can be used as an effective gluon propagator in a $Q(p_1)q(p_2)\rightarrow Q(p_3)q(p_4)$ scattering, leading to a scattering matrix as
\begin{equation}
	\begin{aligned}
		i\mathcal{M} &= \;i\mathcal{M}_\mathrm{Y}+i\mathcal{M}_\mathrm{S}\\
           &= \;\bar{u}^{s'}(p_{3})\gamma^{\mu}u^{s}(p_{1})V_\mathrm{Y}\bar{u}^{r'}(p_{4})\gamma_{\mu}u^{r}(p_{2}) \\
           &+\; \bar{u}^{s'}(p_{3})u^{s}(p_{1})V_\mathrm{S}\bar{u}^{r'}(p_{4})u^{r}(p_{2}).
           \label{eq_iM}
	\end{aligned}
\end{equation}
Following Ref.~\cite{Riek:2010fk}, we assume a vector interaction vertex for the Yukawa term and a scalar interaction vertex for the string term. They do not interfere with each other. The squared matrix element reads
\begin{align}
\label{eq:ScatterAmplitudeSquare}
|\mathcal{M}_{Qq}|^{2}
&=\frac{64\pi^{2}\alpha_\mathrm{s}^{2}}{9}
\frac{(s-m_{Q}^{2})^{2}+(m_{Q}^{2}-u)^{2}+2m_{Q}^{2}t}{(t-m_{d}^{2})^{2}} \nonumber \\
&\quad +\frac{(8\pi\sigma)^{2}}{N_{c}^{2}-1}
\frac{t^{2}-4m_{Q}^{2}t}{(t-m_{s}^{2})^{4}},
\end{align}
with $s$, $t$, $u$ the Mandelstam variables, in which $t=-|\vec{q}|^2$. Since Eq.~(\ref{eq:CornellPotential}) was originally proposed to describe the quark-anti-quark interaction in the color singlet state of heavy quarkonium~\cite{Megias:2007pq}, while other intermediate states can exist in general heavy-light-quark scattering, we apply an overall factor of $1/(N_c^2-1)$ here, with $N_c=C_A=3$, so that the Yukawa part (the first term) reproduces the well-established leading-order (LO) perturbative result of $|\mathcal{M}_{Qq}|^2$~\cite{Combridge:1978kx}. The string part (the second term) introduces the non-perturbative correction to the heavy-light-quark scattering. 

For a heavy-quark-gluon scattering, the string term is introduced only for the \(t\)-channel scattering, with a color factor $C_A/C_F$ introduced relative to the string term in Eq.~(\ref{eq:ScatterAmplitudeSquare}). The full LO perturbative result, including $s$, $t$, and $u$ contributions, is kept for the Yukawa term. The corresponding squared matrix element reads 
\begin{align}
  \label{eq:M2_cg}
  |\mathcal{M}_{Qg}&|^2 =  \frac{64\pi^2 \alpha_\mathrm{s}^2}{9} \frac{(s-m_Q^2)(m_Q^2-u)+2m_Q^2 (s+m_Q^2)}{(s-m_Q^2)^2}\nonumber
  \\+\, & \frac{64\pi^2 \alpha_\mathrm{s}^2}{9} \frac{(s-m_Q^2)(m_Q^2-u)+2m_Q^2 (u+m_Q^2)}{(u-m_Q^2)^2}\nonumber
  \\+\, & 8\pi^2 \alpha_\mathrm{s}^2 \frac{5m_Q^4 + 3m_Q^2t -10m_Q^2u + 4t^2 + 5tu + 5u^2}{(t-m_d^2)^2}\nonumber
  \\+\, & 8\pi^2 \alpha_\mathrm{s}^2 \frac{(m_Q^2-s)(m_Q^2-u)}{(t-m_d^2)^2}\nonumber
  \\+\, & 16\pi^2 \alpha_\mathrm{s}^2 \frac{3m_Q^4 -3m_Q^2s - m_Q^2u + s^2}{(s-m_Q^2)(t-m_d^2)}\nonumber
  \\+\, & \frac{16\pi^2 \alpha_\mathrm{s}^2}{9} \frac{m_Q^2 (4m_Q^2 - t)}{(s - m_Q^2)(m_Q^2 - u)}\nonumber
  \\+\, & 16\pi^2 \alpha_\mathrm{s}^2 \frac{3m_Q^4 - m_Q^2s -3m_Q^2u +u^2}{(t - m_d^2)(u - m_Q^2)}\nonumber
  \\+\, & \frac{C_A}{C_F} \frac{(8\pi \sigma)^2}{N_c^2 -1} \frac{t^2-4m_Q^2 t}{(t-m_s^2)^4}.
\end{align}
The Yukawa terms of Eqs.~(\ref{eq:ScatterAmplitudeSquare}) and~(\ref{eq:M2_cg}) are proportional to $\alpha_\mathrm{s}^2$, in which one $\alpha_\mathrm{s}$ comes from the interaction vertex connecting the medium parton and the exchanged gluon, and the other comes from the vertex connecting the heavy quark and the exchanged gluon. For the former, we use a fixed value $\alpha_\mathrm{s}=0.27$, the same as that used in Eq.~(\ref{eq:CornellPotentialMomentumSpace}); for the latter, we assume it runs with both the heavy quark energy and the medium temperature as $\alpha_\mathrm{s} = 4\pi/[9\ln(2ET/\Lambda_{\rm QCD}^2)]$~\cite{Cao:2017hhk,Xing:2021xwc}. 

Using these matrix elements, the $\hat{q}$ coefficient of a heavy quark can be calculated by
\begin{align}
\label{eq:qhat}
\hat{q}_a (\vec{p}_{a},T) 
&=\frac{\gamma_{b}}{2E_{a}}\sum_{b,d}\int\frac{d^{3}p_{b}}{(2\pi)^{3}2E_{b}}
\frac{d^{3}p_{c}}{(2\pi)^{3}2E_{c}}
\frac{d^{3}p_{d}}{(2\pi)^{3}2E_{d}}\nonumber \\
\phantom{\Gamma_{ab\rightarrow cd}}
 \times\, & f_{b}(E_{b},T)
[1\pm f_{d}(E_{d},T)] \theta(s-(m_{a}+m_{d})^{2})\nonumber \\
\phantom{\Gamma_{ab\rightarrow cd}}
 \times\, & (2\pi)^{4}
\delta^{(4)}(p_{a}+p_{b}-p_{c}-p_{d})
|\mathcal{M}_{ab\rightarrow cd}|^{2}\nonumber \\
 \times\, & [\vec{p}_c-(\vec{p}_c\cdot \hat{p}_a)\hat{p}_a]^2.
\end{align}
Here, the subscripts $a$ and $c$ represent the incoming and outgoing heavy quarks in an elastic $ab\rightarrow cd$ scattering, and the summation runs over all possible scattering channels. The factor $\gamma_b$ denotes the spin-color degeneracy of the medium parton $b$, and thermal distributions (Bose distributions for gluons and Fermi distributions for quarks) are used for $f_{b/d}$ in the local rest frame of the medium. For heavy quark scattering with the QGP medium, we take $m_c=1.27$~GeV for charm quarks and assume zero masses for thermal quarks and gluons inside the medium. As seen in Eq.~(\ref{eq:qhat}), $\hat{q}$ and $\kappa=\hat{q}/2$ depend on both the heavy quark energy and the medium temperature. They control the strengths of medium-induced gluon emissions and quasi-elastic scatterings, respectively, as discussed in the previous subsection.

\begin{figure}[tbp]
    \centering
    \includegraphics[clip=,width=0.4\textwidth]{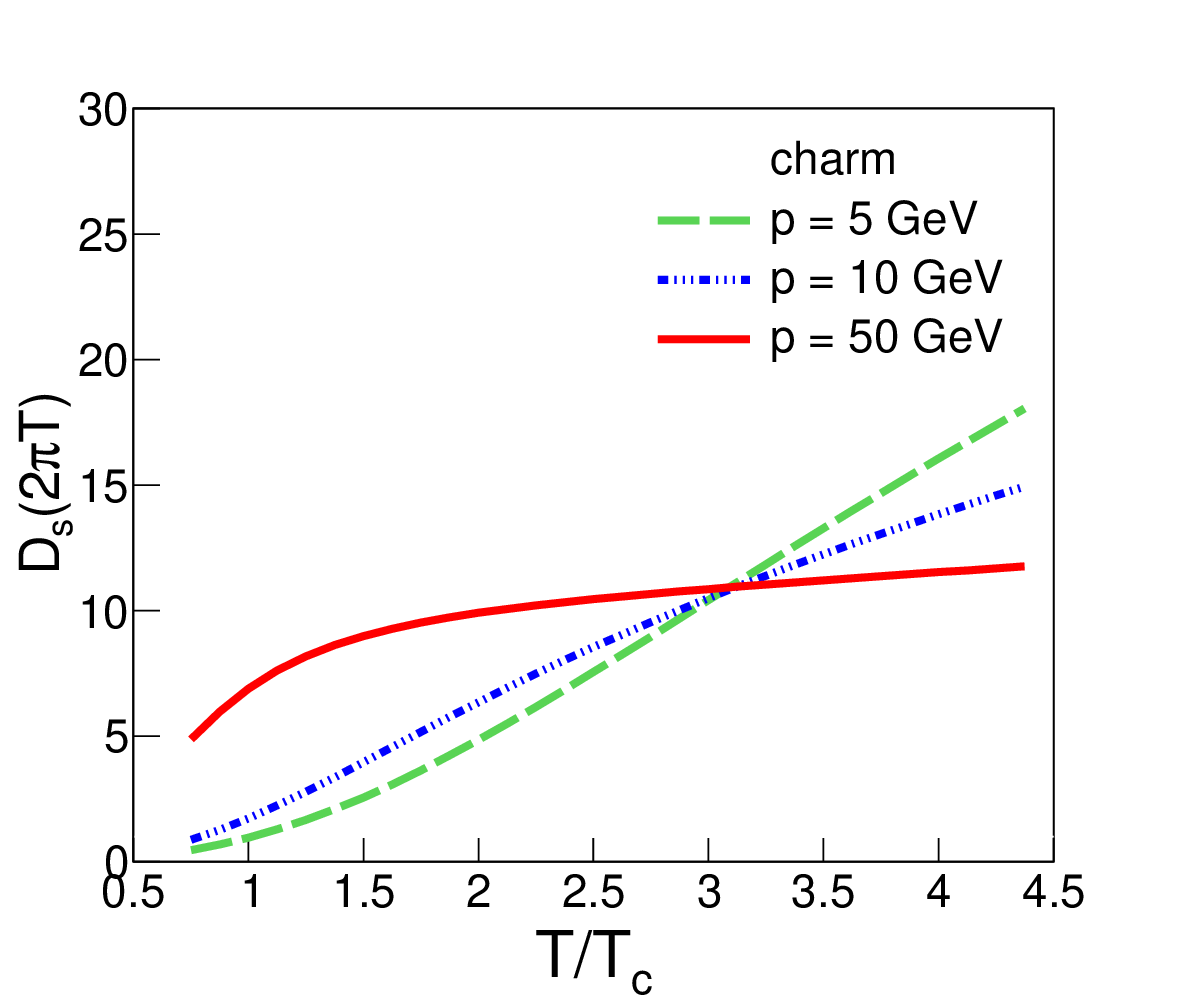}
    \includegraphics[clip=,width=0.4\textwidth]{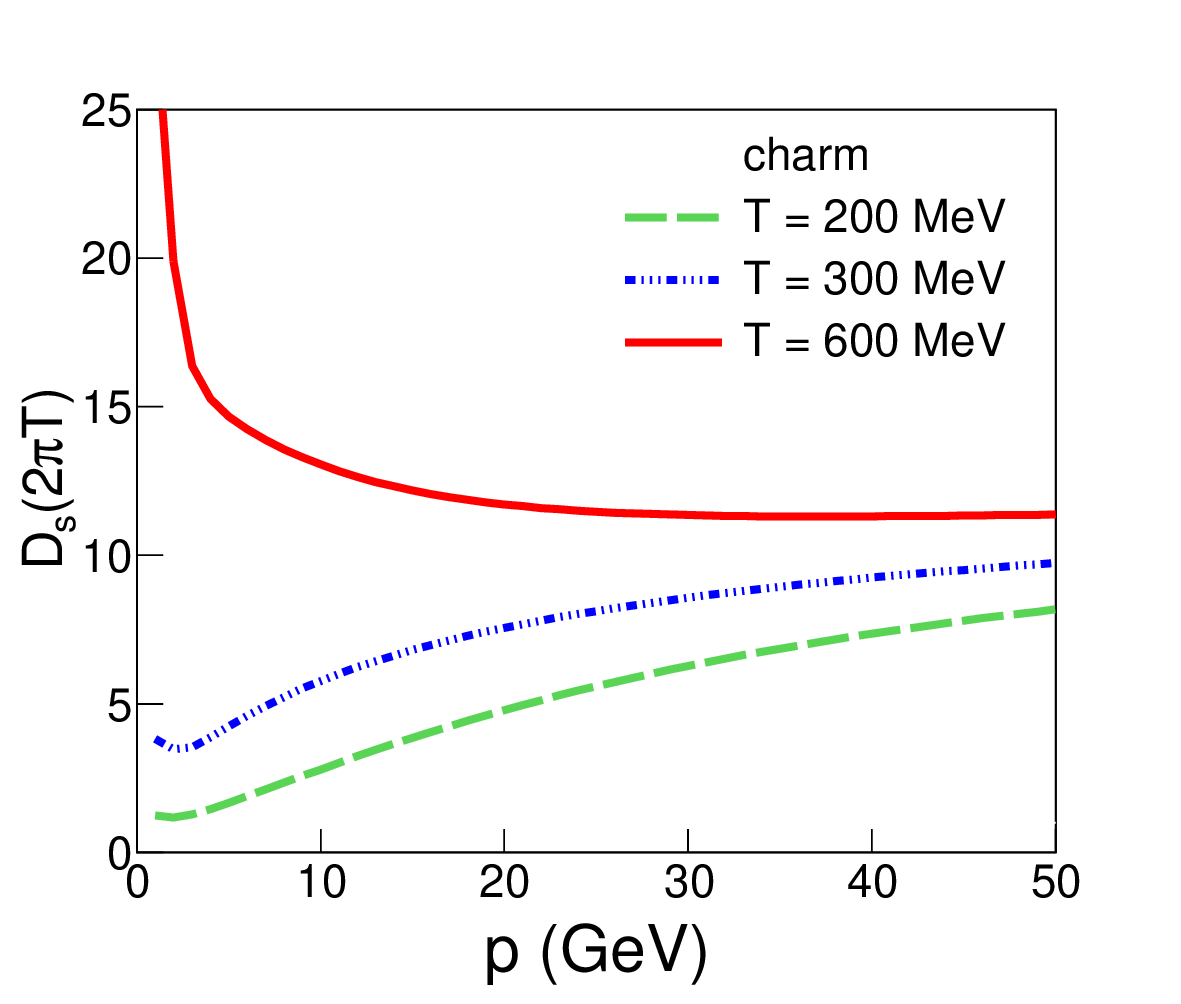}
    \caption{(Color online) The spatial diffusion coefficient of charm quarks \(D_\mathrm{s}(2\pi T)\), upper panel as a function of the medium temperature for different charm quark momenta, lower panel as a function of the charm quark momentum for different medium temperatures.}
    \label{fig:Ds2piT}
\end{figure}

\begin{figure}[tbp]
    \centering
    \includegraphics[clip=,width=0.4\textwidth]{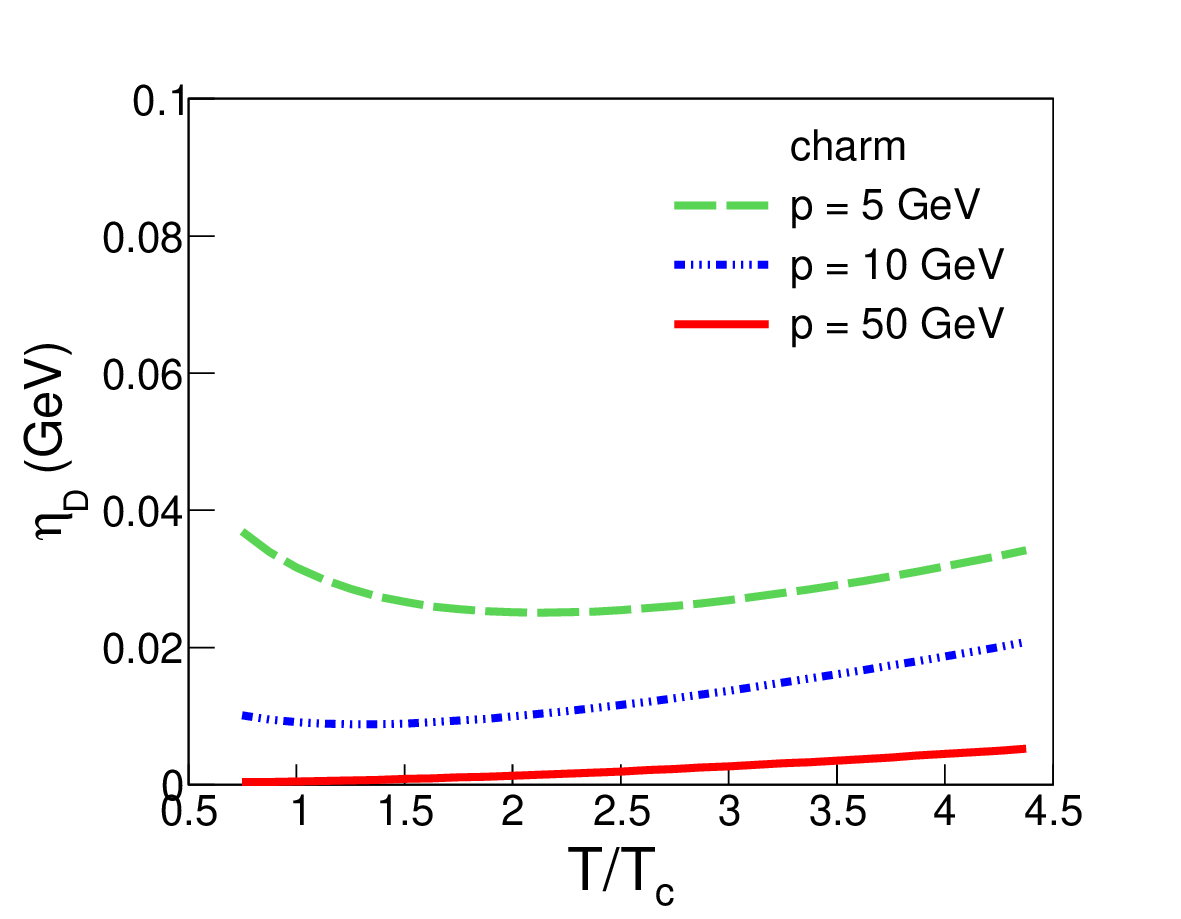}
    \includegraphics[clip=,width=0.4\textwidth]{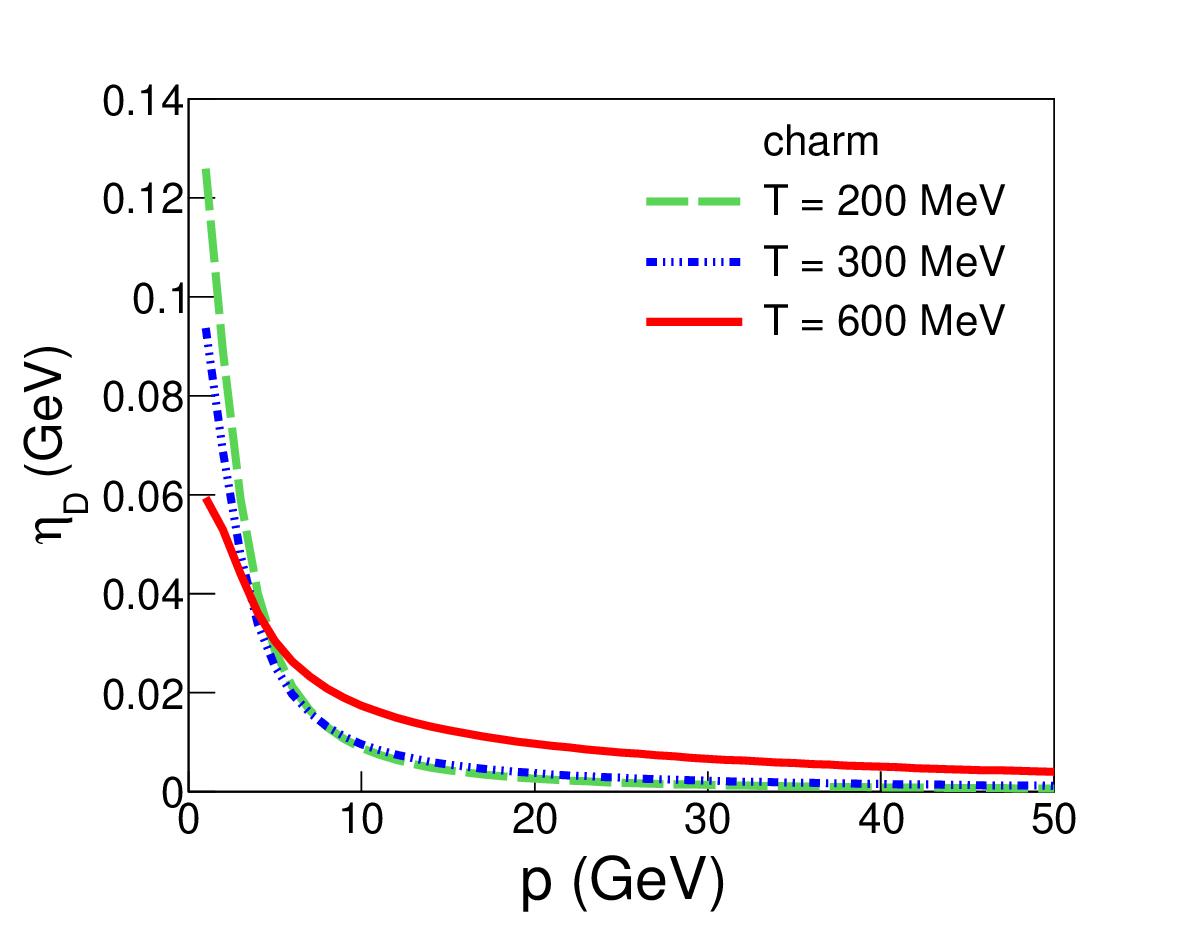}
    \caption{(Color online) The drag coefficient of charm quarks \(\eta_\mathrm{D}\), upper panel as a function of the medium temperature for different charm quark momenta, lower panel as a function of the charm quark momentum for different medium temperatures.}
    \label{fig:dragCoefficient}
\end{figure}

The spatial diffusion coefficient $D_\mathrm{s}$ is a commonly quoted quantity in the literature for measuring the interaction strength between heavy quarks and the QGP medium, a smaller $D_\mathrm{s}$ corresponding to a stronger interaction. We convert $\kappa$ into $D_\mathrm{s}$ using Eqs.~(\ref{eq:generalizedFDT}) and~(\ref{eq:SpatialDiffusionCoefficient}), and present the dimensionless quantity \(D_\mathrm{s}(2\pi T)\) in Fig.~\ref{fig:Ds2piT}: the upper panel shows its temperature dependence for three different charm quark momenta (5, 10, and 50~GeV), and the lower panel shows its momentum dependence for three different medium temperatures (200, 300, and 600~MeV). The pseudo-critical temperature for hadronization, $T_\mathrm{c}=160$~MeV, is taken as a reference temperature in the figure. In the upper panel, we can see that \(D_\mathrm{s}(2\pi T)\) increases with the medium temperature, and the slope of the increase is smaller for charm quarks with larger momenta. This can be understood with the weaker string interaction, or non-perturbative effect, at higher charm quark momenta and higher medium temperatures. It is conventional to compare \(D_\mathrm{s}(2\pi T)\) at zero momentum to the lattice QCD result, although its value at finite momentum is more relevant for heavy quark phenomenology in heavy-ion collisions. As shown in a separate work~\cite{Zhang:2025kjr}, the temperature dependence of this \(D_\mathrm{s}(2\pi T)\) in the zero momentum limit from our calculation is consistent with the recent lattice QCD data~\cite{HotQCD:2025fbd} and other model calculations that take into account non-perturbative effects~\cite{Liu:2016ysz,Tang:2023tkm,Sambataro:2024mkr}. In the lower panel, we see that at very high temperature, \(D_\mathrm{s}(2\pi T)\) decreases and then becomes flat as the charm quark momentum becomes larger; while at low temperature, \(D_\mathrm{s}(2\pi T)\) increases with the charm quark momentum. This results from the competition between the Yukawa and string interactions. Under the Yukawa interaction, which dominates at high temperature, $\kappa$ first increases with the charm quark momentum and then becomes nearly momentum independent at high momentum. On the other hand, the string interaction, which dominates at low temperature, becomes weaker as the charm quark momentum increases. The diffusion coefficient here has markedly different dependences on temperature and momentum from the simple parameterizations used in our earlier work, which will affect the heavy flavor observables obtained from our model calculations.

In Fig.~\ref{fig:dragCoefficient}, we present the drag coefficient \(\eta_\mathrm{D}\) of charm quarks, which quantifies the speed of heavy quark energy (or longitudinal momentum) loss inside the QGP. Although $\hat{q}/T^3$ (or $\kappa/T^3$) decreases as the medium temperature increases~\cite{Xing:2021xwc}, $\hat{q}$ (or $\kappa$) itself generally increases with temperature since the density of scattering centers inside the medium increases rapidly ($\sim T^3$) with temperature. Similarly, in general, \(\eta_\mathrm{D}\) increases with the medium temperature, as seen in the upper panel of Fig.~\ref{fig:dragCoefficient} for charm quarks with large momenta. However, at low momentum, string interaction becomes strong and significantly enhances \(\eta_\mathrm{D}\) at low temperature, resulting in its non-monotonic temperature dependence for low momentum charm quarks. In the lower panel of Fig.~\ref{fig:dragCoefficient}, we show the momentum dependence of $\eta_\mathrm{D}$. When both Yukawa and string interactions are included, $\hat{q}$ (or $\kappa$) first decreases and then varies slowly as the charm quark momentum increases~\cite{Xing:2021xwc}. With the additional factor of energy in the denominator when $\kappa$ is converted into $\eta_\mathrm{D}$, Eq.~(\ref{eq:generalizedFDT}), $\eta_\mathrm{D}$ decreases with the charm quark momentum.

\subsection{Thermalization of charm quarks in a static medium}
\label{sec:Thermalization}

\begin{figure*}[tbp]
    \centering
    \includegraphics[clip=,width=0.346\textwidth]{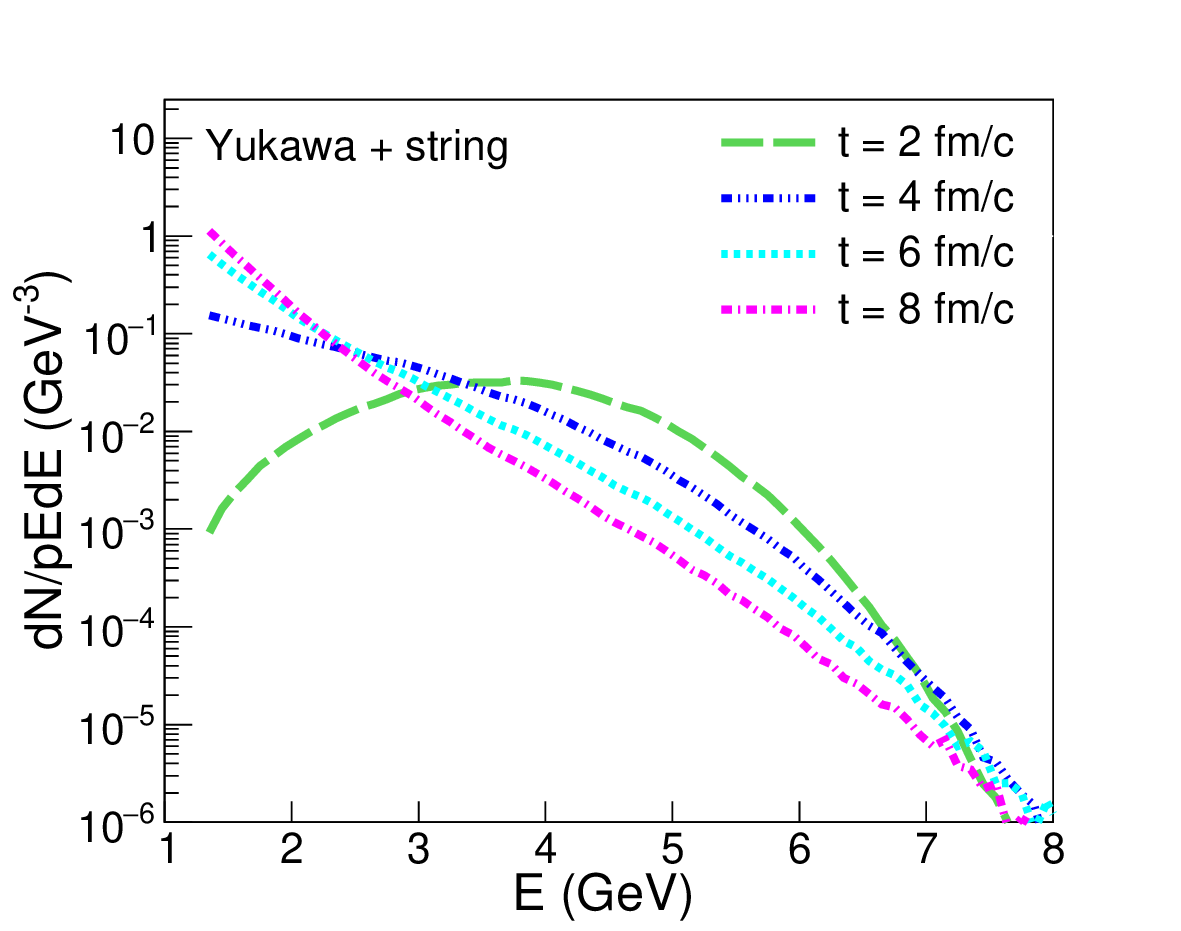}
    \hspace{-20pt}
    \includegraphics[clip=,width=0.346\textwidth]{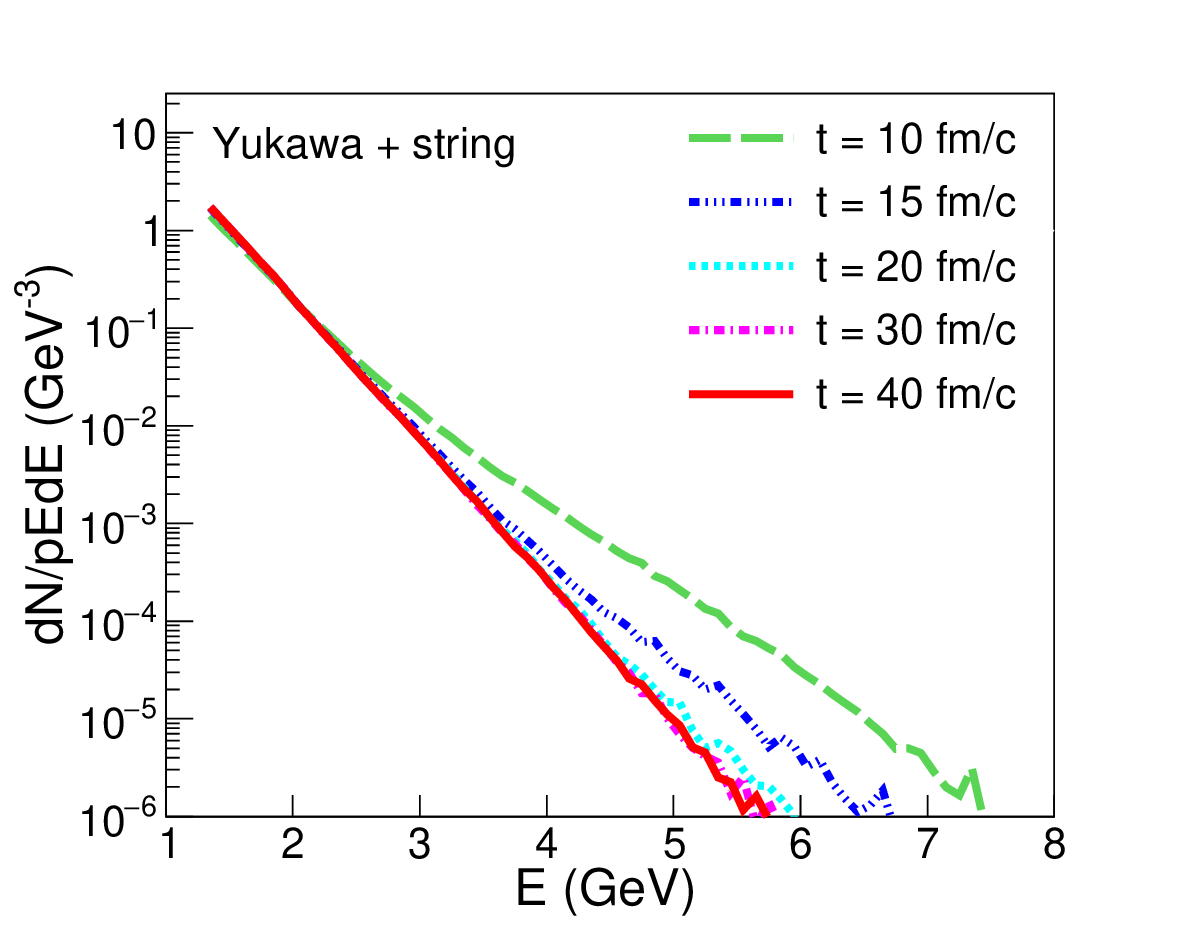}
    \hspace{-20pt}
    \includegraphics[clip=,width=0.35\textwidth]{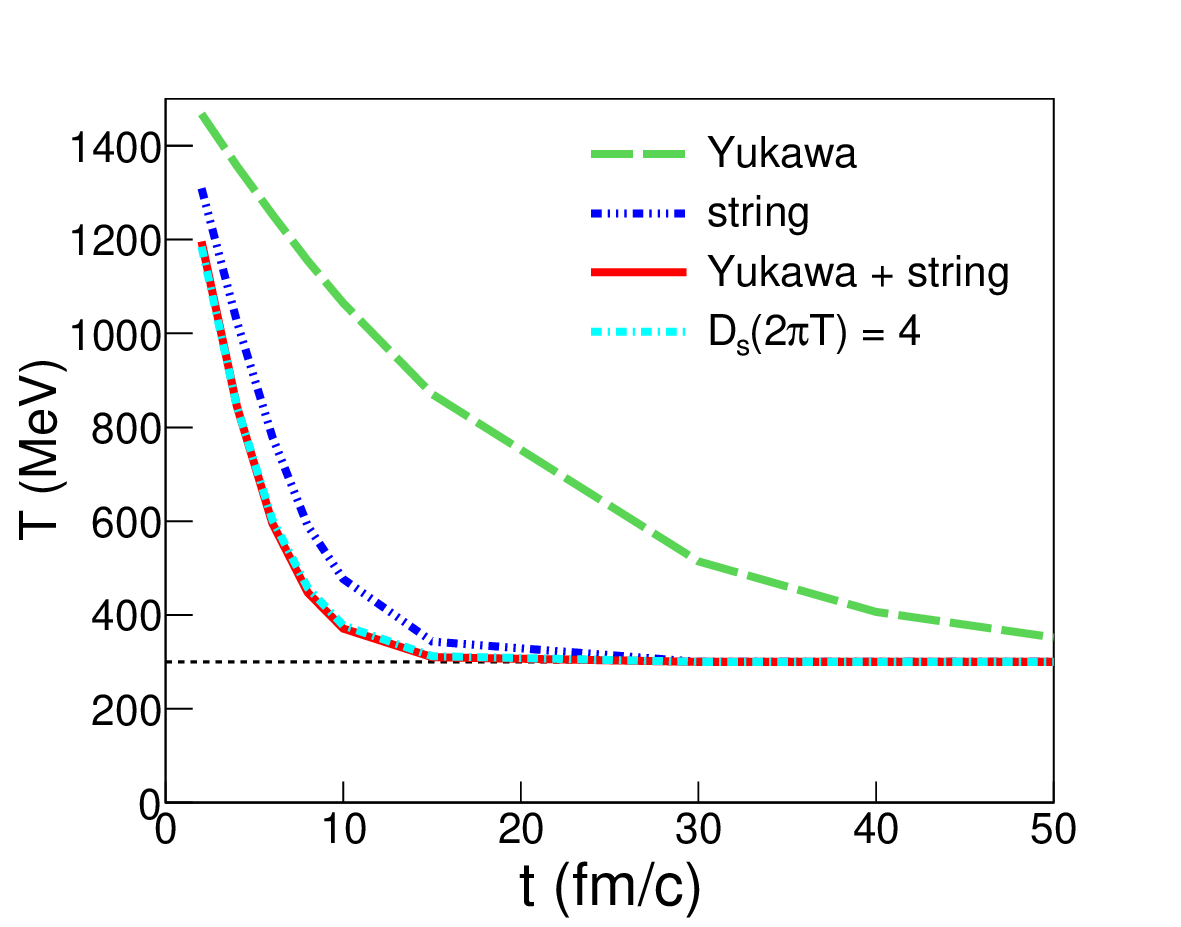}
    \caption{(Color online) Left two panels: the charm quark energy spectra at different times, starting as a $\delta$-function at $p=5$~GeV and evolving under Yukawa+string interactions inside a static medium at $T=300$~MeV. Right panel: time evolution of the temperature parameter of the charm quark system, extracted from a Boltzmann fit to the charm quark spectrum, and compared between different model setups: Yukawa interaction, string interaction, Yukawa+string interaction, and a constant \(D_s(2\pi T)=4\).}
    \label{fig:thermalization}
\end{figure*}

As a crucial validation of the Langevin model, we examine the thermal equilibrium limit of the heavy quark evolution inside a static medium at a fixed temperature. For heavy quarks that reach thermal equilibrium with the medium, their energy spectrum can be approximated by the following Boltzmann distribution~\cite{Cao:2011et}
\begin{equation}
\label{eq:HQMomentumSpectra-3}
\frac{dN}{pEdE}=Ce^{-E/T},
\end{equation}
with $p$ and $E$ the magnitude of momentum and energy of heavy quarks, respectively, and $C$ a constant. Therefore, we may analyze the energy spectrum of heavy quarks, use the expression above to extract the temperature parameter ($T$) of the heavy quark system, and compare it to the medium temperature as time evolves.

We initialize charm quarks with a fixed momentum of 5~GeV along a given ($\hat{z}$) direction and let them evolve through a static and infinite medium at $T=300$~MeV. Shown in Fig.~\ref{fig:thermalization} are the energy spectra and the extracted temperature parameters of the charm quark systems at different times. In the left two panels, both the Yukawa and string interactions are included in charm quark scatterings with the medium, and each spectrum is normalized to one charm quark. One can observe that at early time, the charm quark spectrum quickly evolves from a $\delta$-function at $p=5$~GeV towards a decreasing function starting from $p=0$ (or $E=m_c$). A Boltzmann-like function (linear decrease on the log scale) occurs at $t\sim 10$~fm/$c$, though the (absolute value of the) slope continues to increase until $t\sim 20$~fm/$c$, beyond which the charm quark spectrum stays almost still, signifying the thermal equilibrium of the charm quark system with its surrounding medium. In the right panel, we extract the temperature parameter of the charm quark system using the Boltzmann fit to the charm quark spectrum. Besides including both the Yukawa and string interactions, we present results for their individual contributions and for a constant diffusion coefficient at $D_\mathrm{s}(2\pi T)=4$. Since the string interaction is much stronger than the Yukawa interaction for low momentum charm quarks, the former dominates the thermalization process of the system. When they are combined, the charm quark system approaches the medium temperature at $t\sim 20$~fm/$c$, consistent with the features of the charm quark spectra in the middle panel. Our improved Langevin model is able to produce the correct thermal equilibrium limit of the system. Notably, the current model, which takes into account the energy and momentum dependence of the diffusion coefficients, has a thermalization speed comparable to that of the previous version of our model, which used a constant $D_\mathrm{s}(2\pi T)=4$. This indicates that the overall interaction strength between heavy quarks and the medium is similar in the two versions after they are constrained by the heavy flavor data from experiments.


\section{heavy flavor hadron observables in heavy-ion collisions}
\label{sec:results}

To study the heavy flavor phenomenology in heavy-ion collisions, we embed the improved Langevin model in the following framework for numerical simulation. The Monte-Carlo Glauber model is used to generate the initial spatial distributions of both heavy quarks and the energy density of the QGP medium. The momentum space of heavy quarks is initialized with the fixed-order-next-to-leading-log (FONLL) calculation~\cite{Cacciari:2001td,Cacciari:2012ny,Cacciari:2015fta} convoluted with the CT14NLO~\cite{Dulat:2015mca} parton distribution function. The EPPS16~\cite{Eskola:2016oht} parametrization at the next-to-leading order is applied for the nuclear shadowing effect in nucleus-nucleus collisions. Heavy quarks are assumed to stream freely before the hydrodynamic stage of the QGP medium (\(\tau_{0}=0.6~\mathrm{fm}\)/$c$). After that, we use the Langevin model to simulate the heavy quark interaction with the QGP medium. The QGP medium is generated using the \((3+1)\)-dimensional viscous hydrodynamic model CLVisc~\cite{Pang:2012he,Pang:2014ipa,Wu:2018cpc,Wu:2021fjf}. The hydrodynamic simulation provides the spacetime profile of the QGP fireball, including its local temperature and flow velocity. Using this information, at each time step, we first boost each heavy quark into the local rest frame of the medium, in which its momentum is updated according to the Langevin equation. The heavy quark is then boosted back to the global frame and streams freely for one time step. This process is iterated until the heavy quark exits the color-deconfined region of the nuclear matter. On the hadronization hypersurface, defined with \(T_\mathrm{c}=160~\mathrm{MeV}\), we convert heavy quarks into heavy flavor hadrons through a hybrid fragmentation-coalescence model~\cite{Cao:2019iqs}. The coalescence module is based on the instantaneous coalescence picture, in which the probability of coalescence is determined by the wavefunction overlap between the free quark state and the hadron bound states. Both \(s\)- and \(p\)-wave states of charm hadrons are taken into account in the coalescence process. For heavy quarks that do not form hadrons via coalescence, we convert them to hadrons via the fragmentation module of Pythia. In this section, we focus on how different interaction mechanisms, Yukawa, string, and our earlier assumption of constant $D_\mathrm{s} (2\pi T)$, affect the medium-modified momentum spectrum of charm quarks and therefore the charm hadron observables in heavy-ion collisions.

\subsection{Heavy meson suppression and flow at RHIC and the LHC}
\label{sec:raa-v2}

\begin{figure}[tbp]
    \centering
    \includegraphics[clip=,width=0.4\textwidth]{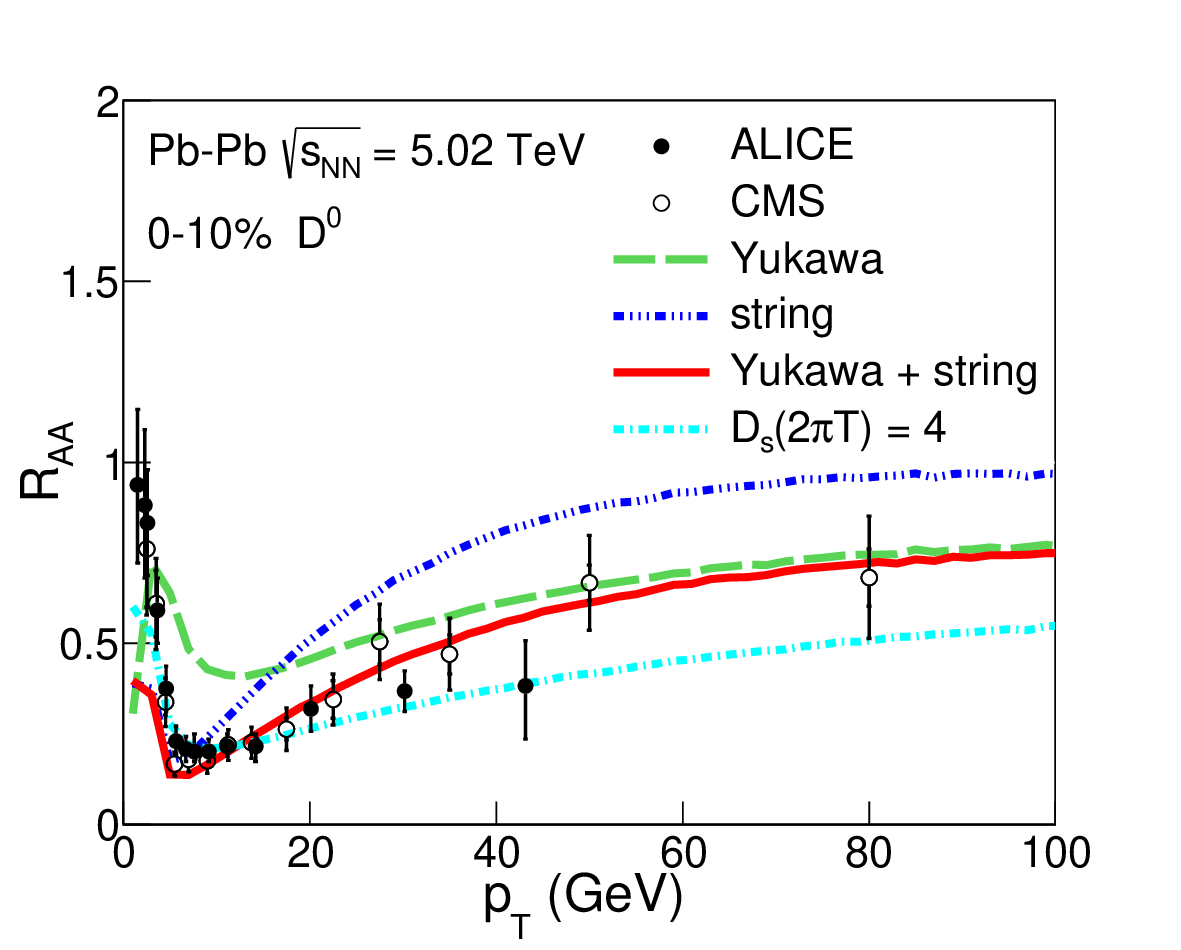}
    \includegraphics[clip=,width=0.4\textwidth]{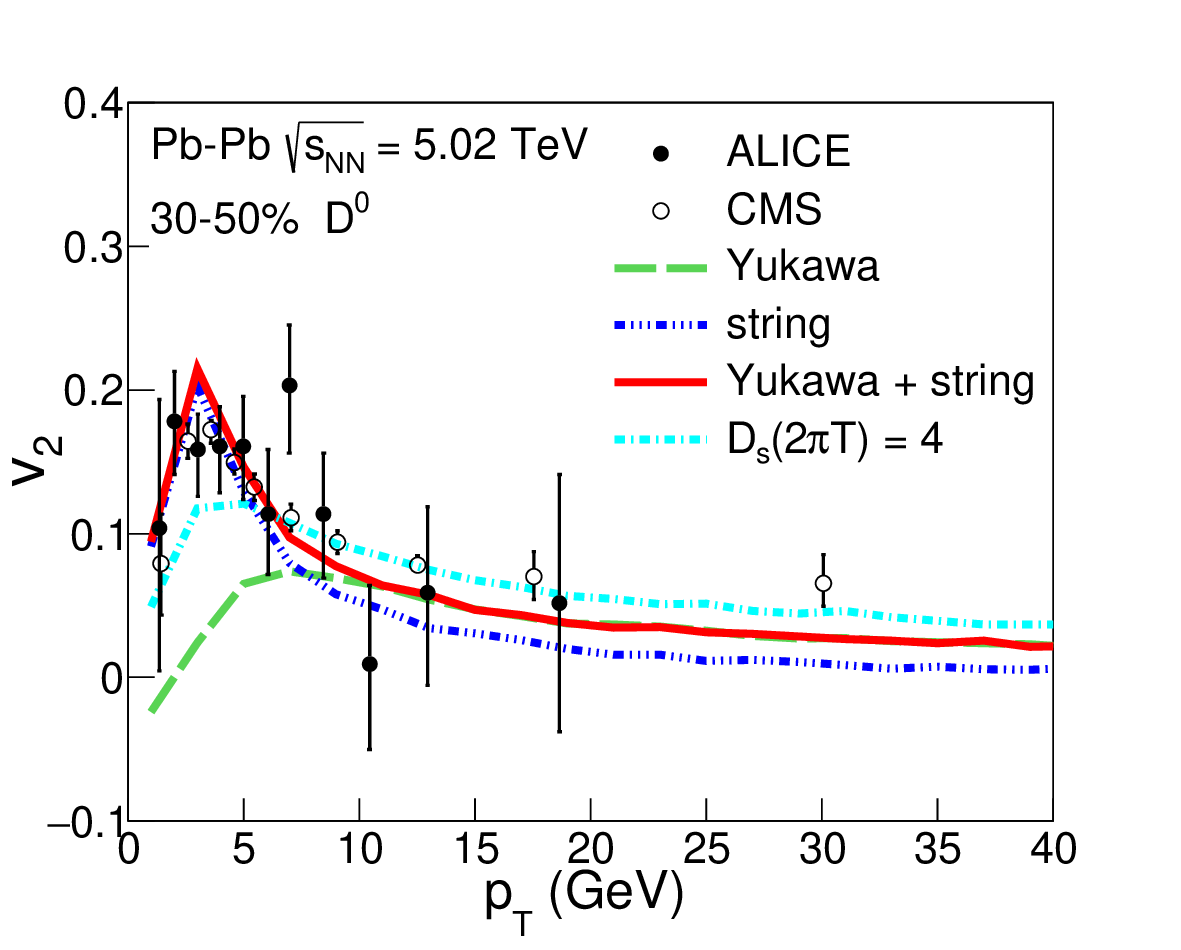}
    \caption{(Color online) The nuclear modification factor \(R_\mathrm{AA}\) (upper panel) and elliptic flow coefficient \(v_2\) (lower panel) of \(D\) mesons in Pb+Pb collisions at \(\sqrt{s_{\mathrm{NN}}}=5.02~\mathrm{TeV}\), compared between contributions from the Yukawa interaction, string interaction, their combination, and a constant diffusion coefficient \(D_s(2\pi T)=4\). The CMS~\cite{Sirunyan:2017plt,Sirunyan:2017xss} and ALICE~\cite{Acharya:2017qps,Acharya:2018hre} data are shown for comparison.}
    \label{fig:raa-v2-LHC}
\end{figure}

We first present the \(R_\mathrm{AA}\) and \(v_{2}\) of \(D\) mesons in Pb+Pb collisions at \(\sqrt{s_{\mathrm{NN}}}=5.02~\mathrm{TeV}\) in Fig.~\ref{fig:raa-v2-LHC}. Contributions from the short-range Yukawa and long-range string interactions are investigated separately, and the CMS~\cite{Sirunyan:2017plt,Sirunyan:2017xss} and ALICE~\cite{Acharya:2017qps,Acharya:2018hre} data are shown for comparison. One can clearly observe that the string interaction dominates the heavy quark energy loss at low \(p_\mathrm{T}\), while the Yukawa interaction dominates at high \(p_\mathrm{T}\). Compared to the Yukawa interaction, the string interaction becomes much stronger when the temperature decreases towards the hadronization temperature \(T_\mathrm{c}\) and therefore helps low-to-intermediate-$p_\mathrm{T}$ heavy quarks obtain a larger \(v_{2}\) from the late stage of the QGP when the strong collectivity of the QGP has developed. The heavy meson $v_2$ is driven by different factors at different $p_\mathrm{T}$: at low $p_\mathrm{T}$, it is dominated by the coupling of heavy quarks to the collective motion of the expanding medium; at intermediate $p_\mathrm{T}$, it is affected by the coalescence between heavy and light quarks in the hadronization process, which also relies on the medium collectivity; and at high $p_\mathrm{T}$, it is dominated by the energy loss of heavy quarks through the geometrically asymmetric medium. By combining the Yukawa and string interactions, our improved Langevin model provides a reasonable description of the \(D\) meson \(R_\mathrm{AA}\) and \(v_{2}\) at the LHC. Due to more realistic temperature and momentum dependence of the heavy quark transport coefficient introduced in our current model, a better performance is achieved compared to the earlier version of the model in which a constant \(D_s(2\pi T)=4\) is used. In particular, the current model provides a larger $v_2$ of $D$ mesons at low to intermediate $p_\mathrm{T}$, which is favored by the experimental data.

\begin{figure}[tbp]
    \centering
    \includegraphics[clip=,width=0.4\textwidth]{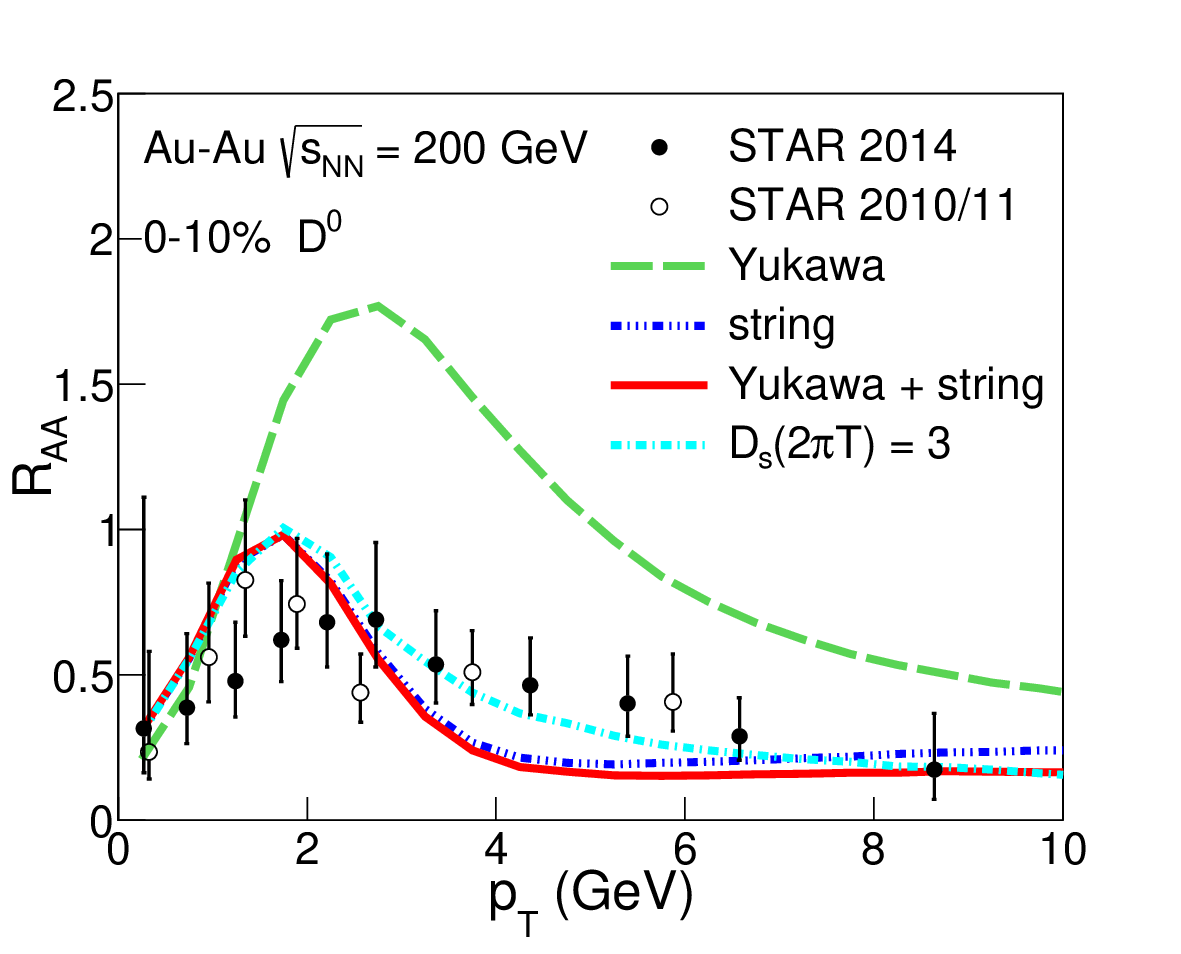}
    \includegraphics[clip=,width=0.4\textwidth]{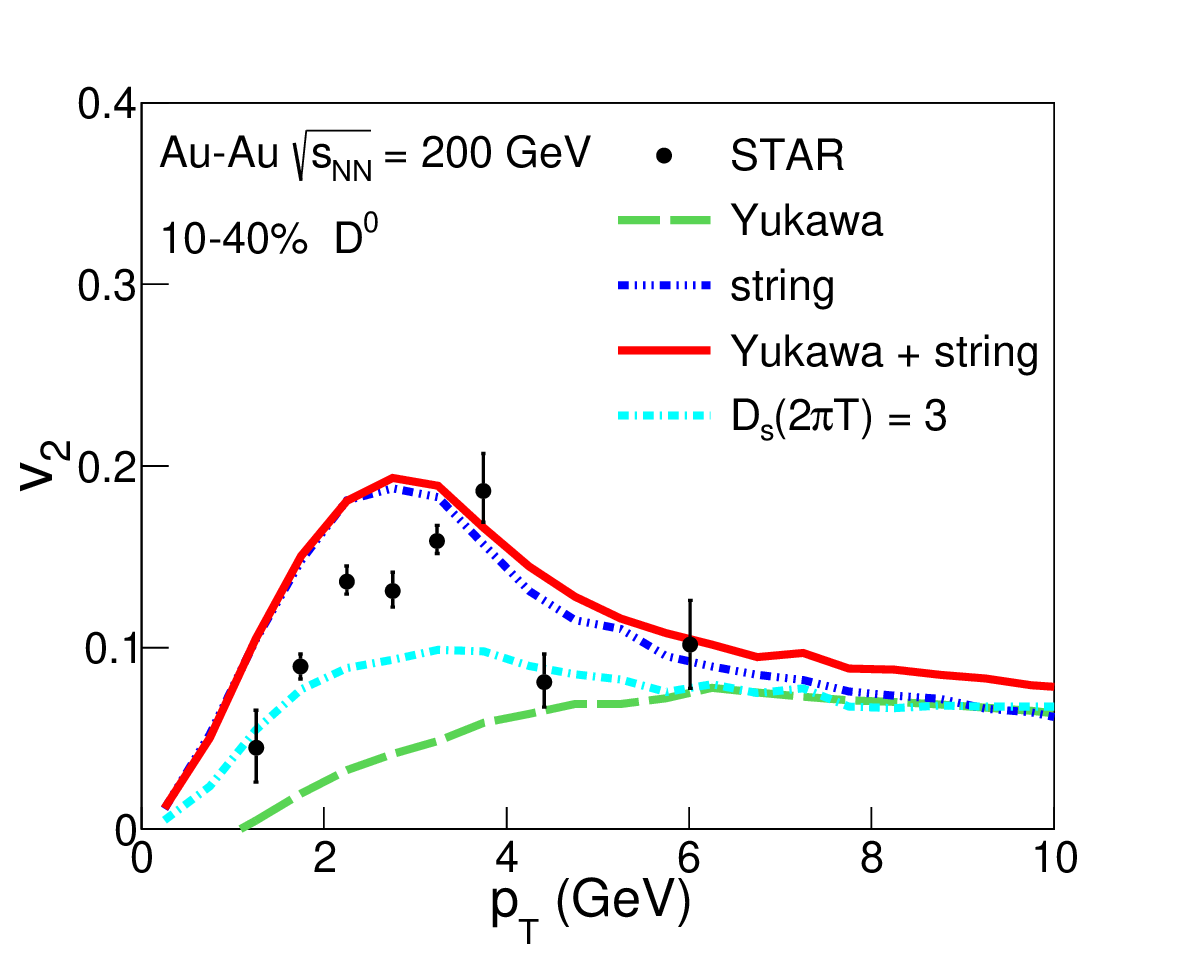}
    \caption{(Color online) The nuclear modification factor \(R_\mathrm{AA}\) (upper panel) and elliptic flow coefficient \(v_2\) (lower panel) of \(D\) mesons in Au+Au collisions at \(\sqrt{s_{\mathrm{NN}}}=200~\mathrm{GeV}\), compared between contributions from the Yukawa interaction, string interaction, their combination, and a constant diffusion coefficient \(D_s(2\pi T)=3\). The STAR data~\cite{Adamczyk:2014uip,Adamczyk:2017xur,Adam:2018inb} are shown for comparison.}
    \label{fig:raa-v2-RHIC}
\end{figure}

\begin{figure*}[tbp]
    \centering
    \includegraphics[clip=,width=0.34\textwidth]{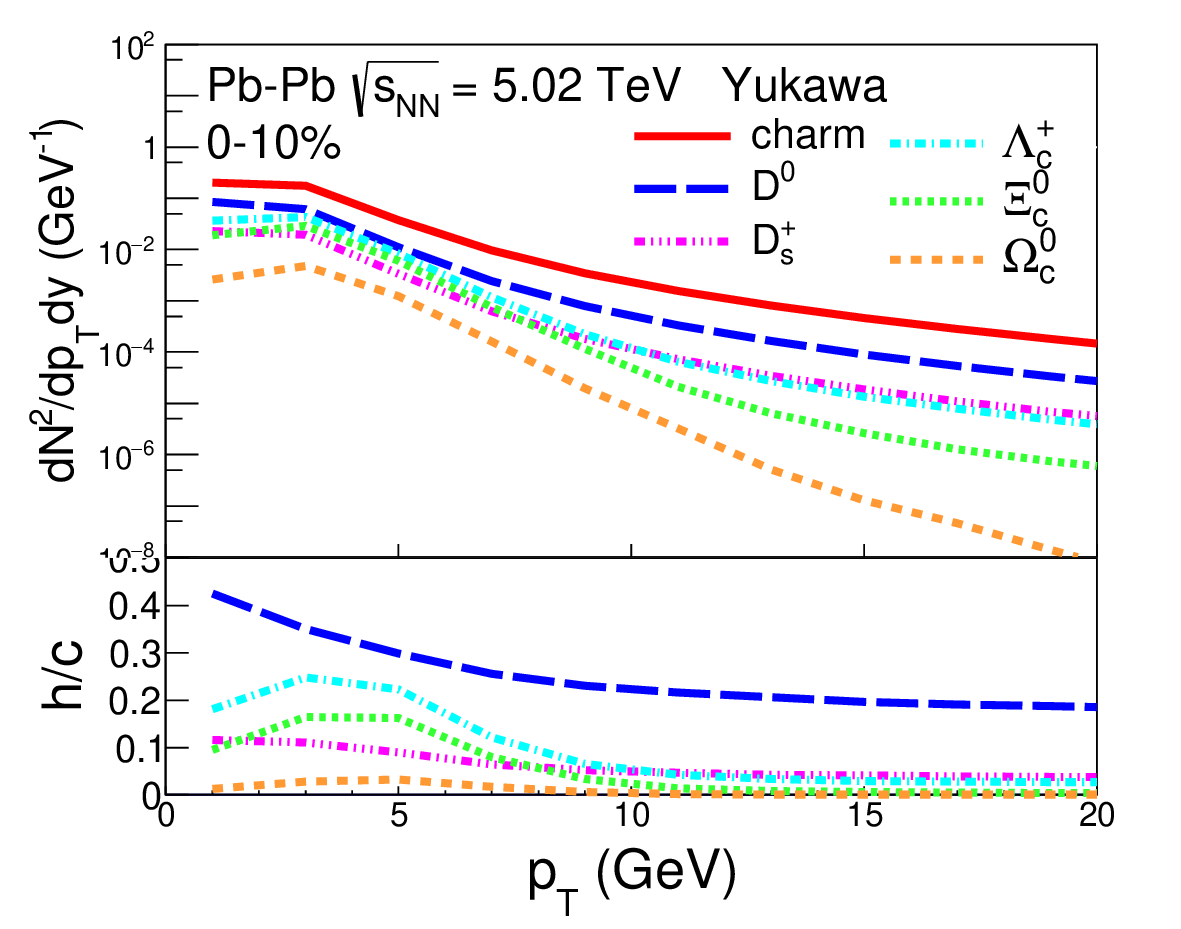}
    \hspace{-14pt}
    \includegraphics[clip=,width=0.34\textwidth]{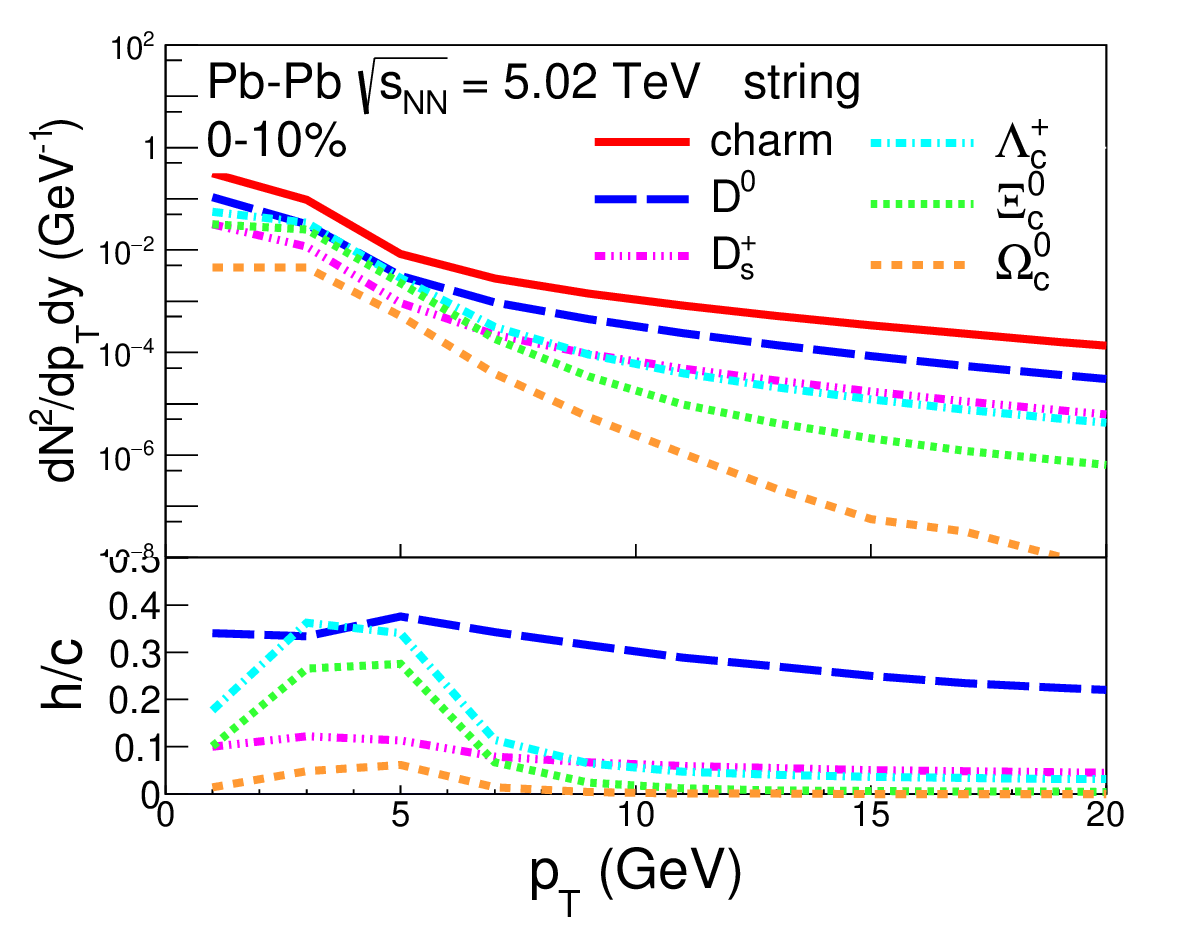}
    \hspace{-14pt}
    \includegraphics[clip=,width=0.34\textwidth]{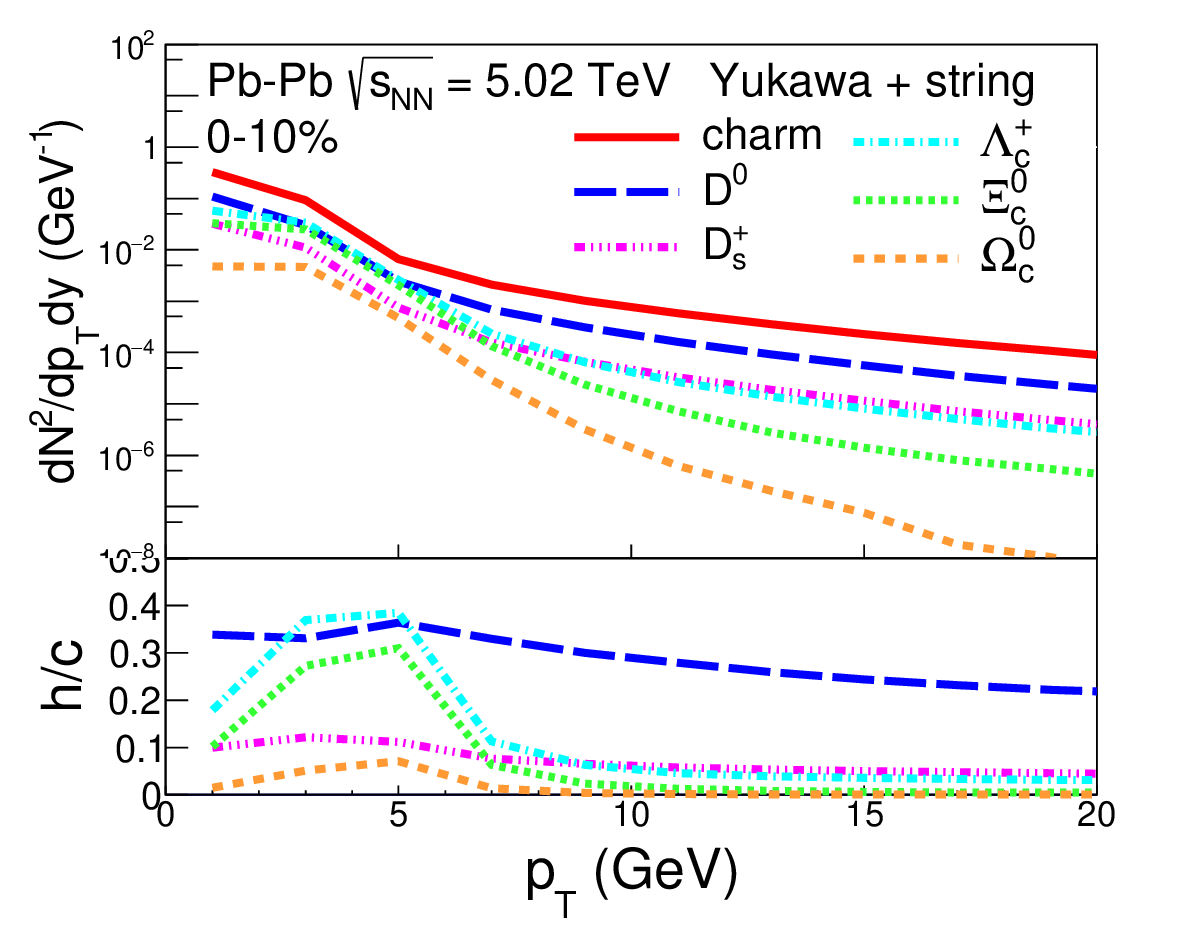}
    \caption{(Color online) The transverse momentum spectra of charm quarks and different species of charm hadrons (upper row) and the hadron-to-quark (\(h/c\)) ratios (lower row) in 0-10\% Pb+Pb collisions at \(\sqrt{s_{\mathrm{NN}}}=5.02~\mathrm{TeV}\), compared between contributions from the Yukawa interaction (left panel), string interaction (middle panel), and their sum (right panel).}
    \label{fig:charm-spectra-ratio}
\end{figure*}

Figure~\ref{fig:raa-v2-RHIC} shows the \(R_\mathrm{AA}\) and \(v_{2}\) of \(D\) mesons in Au+Au collisions at \(\sqrt{s_{\mathrm{NN}}}=200~\mathrm{GeV}\), compared to the STAR data~\cite{Adamczyk:2014uip,Adamczyk:2017xur,Adam:2018inb}. Since the STAR measurement focuses on the low to intermediate $p_\mathrm{T}$ regime, the \(D\) meson observables here are dominated by the string interaction. The  Yukawa interaction alone would significantly overestimate the \(D\) meson \(R_\mathrm{AA}\) and underestimate the \(v_{2}\). After introducing the string interaction, our model provides a reasonable description of the experimental data at the RHIC energy. When a constant \(D_s(2\pi T)\) is used in the Langevin model, one usually needs a smaller value for the RHIC environment than the LHC environment, considering that the average QGP temperature at RHIC is lower than that at the LHC. In Ref.~\cite{Li:2020kax}, \(D_s(2\pi T)=3\) and 4 are tuned for the RHIC and LHC data, respectively, when heavy quark interaction with the pre-equilibrium medium (before $\tau_0=0.6$~fm/$c$) is neglected. In contrast, using the transport coefficient evaluated from the Yukawa-string potential scattering model, our Langevin model is able to describe the RHIC and LHC data simultaneously. Using the same transport coefficient, results from the Langevin model here are also consistent with those obtained from the LBT model~\cite{Xing:2021xwc}.

\subsection{Charm hadron chemistry}
\label{sec:spectra-ratio}

The heavy flavor hadron chemistry, or the ratio between spectra of different heavy flavor hadrons, provides a unique opportunity to probe the hadronization process of free quarks. It is now commonly agreed that the coalescence between heavy quarks and thermal partons from the QGP is essential for understanding the enhanced baryon-to-meson ratio observed in heavy-ion collisions relative to proton-proton and electron-positron collisions. Effects of hadronization on heavy flavor observables and systematic comparisons between different hadronization models have been extensively investigated in the literature~\cite{Rapp:2018qla,Zhao:2023nrz}. On the other hand, within a given hadronization model, the heavy flavor hadron chemistry also strongly depends on the heavy quark spectrum. As discussed in Ref.~\cite{Cao:2019iqs}, the softer heavy quark spectrum in Au+Au collisions at $\sqrt{s_\mathrm{NN}}=200$~GeV results in a larger $\Lambda_c/D^0$ ratio than that in Pb+Pb collisions at $\sqrt{s_\mathrm{NN}}=5.02$~TeV, although the stronger radial flow of the QGP in the latter system is also expected to enhance this ratio. In this subsection, we investigate how different medium-modified spectra of charm quarks affect their hadron chemistry.

Figure~\ref{fig:charm-spectra-ratio} shows the transverse momentum spectra of charm quarks and different species of charm hadrons, together with the hadron-to-quark ratios, in central  (0-10\%) Pb+Pb collisions at \(\sqrt{s_{\mathrm{NN}}}=5.02~\mathrm{TeV}\). Each spectrum is normalized to one charm quark. Comparing the contributions from the Yukawa (left panel) and string (middle panel) interactions, we see the latter results in a stronger accumulation of charm quarks at low $p_\mathrm{T}$ and a more rapid decrease of the medium-modified charm quark spectrum below $p_\mathrm{T}\sim 10$~GeV. This is due to the stronger energy loss of charm quarks caused by the string interaction than by the Yukawa interaction at low $p_\mathrm{T}$. Quarks with closer velocities are easier to combine into hadrons within our coalescence model. Therefore, it is easier for charm quarks closer to thermal equilibrium with the medium to combine with the medium partons into hadrons. More abundant charm quarks at low $p_\mathrm{T}$ lead to a more prominent enhancement of the charm hadrons at the intermediate $p_\mathrm{T}$ (2--6~GeV) via heavy-light-quark coalescence, giving rise to the bump structures of the hadron-to-charm ratios in this regime. Considering that the QGP medium is richer in strangeness and baryon number than string breaking (fragmentation) in vacuum, enhancement of the coalescence between charm quarks and the medium partons also increases the strange-to-non-strange hadron ratios and the baryon-to-meson ratios. Our full result (right panel) is dominated by the string interaction (middle panel) in the low to intermediate $p_\mathrm{T}$ region.

\begin{figure*}[tbp]
    \centering
    \includegraphics[clip=,width=0.35\textwidth]{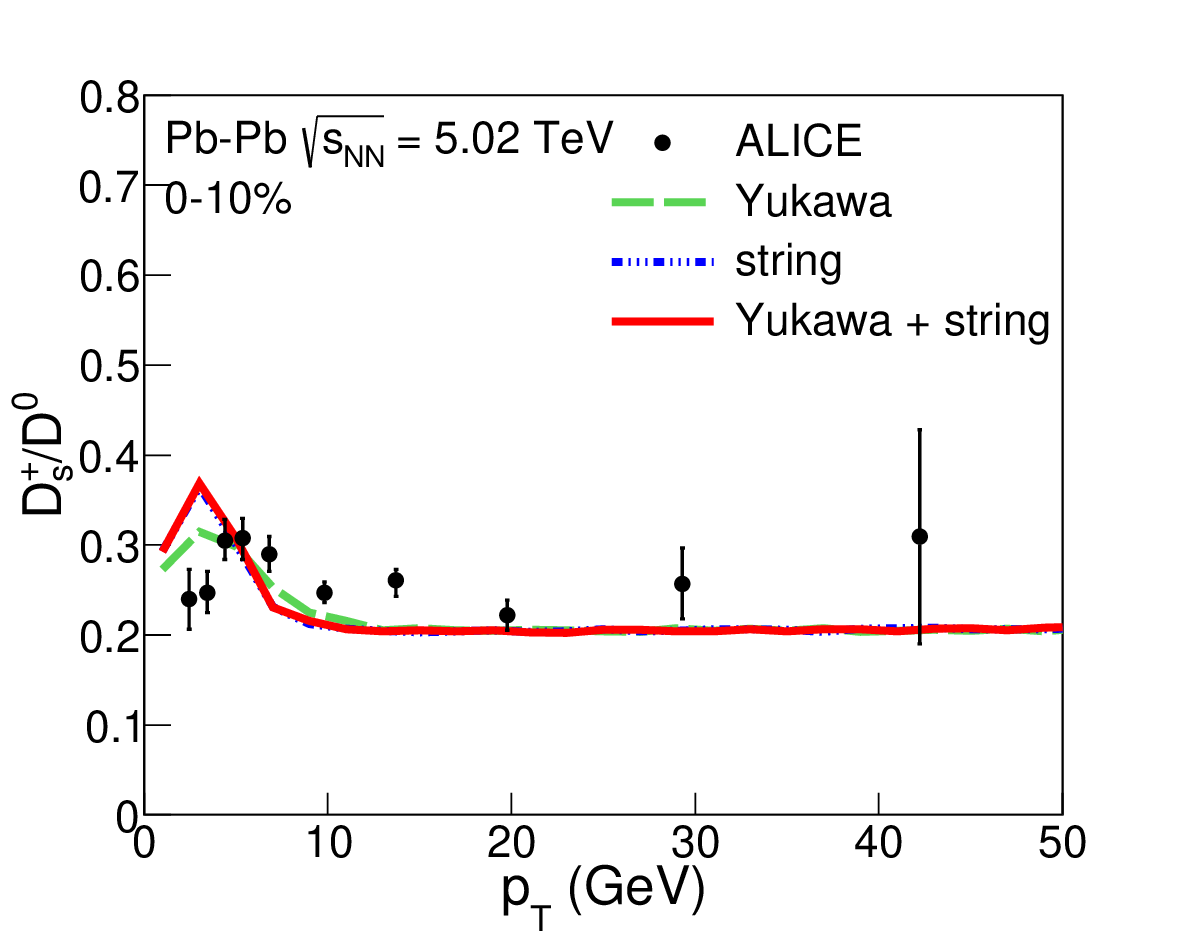}
    \hspace{-20pt}
    \includegraphics[clip=,width=0.35\textwidth]{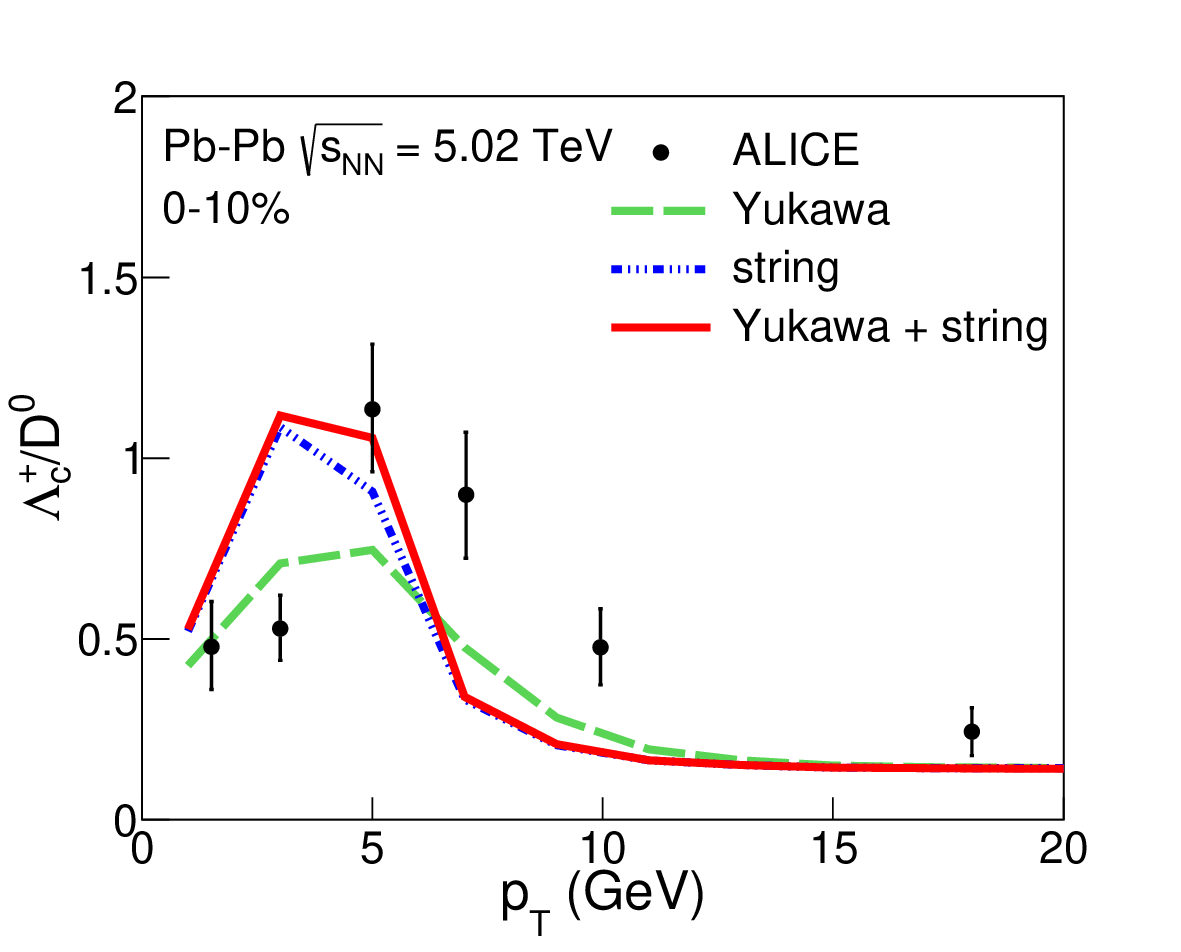}
    \hspace{-20pt}
    \includegraphics[clip=,width=0.35\textwidth]{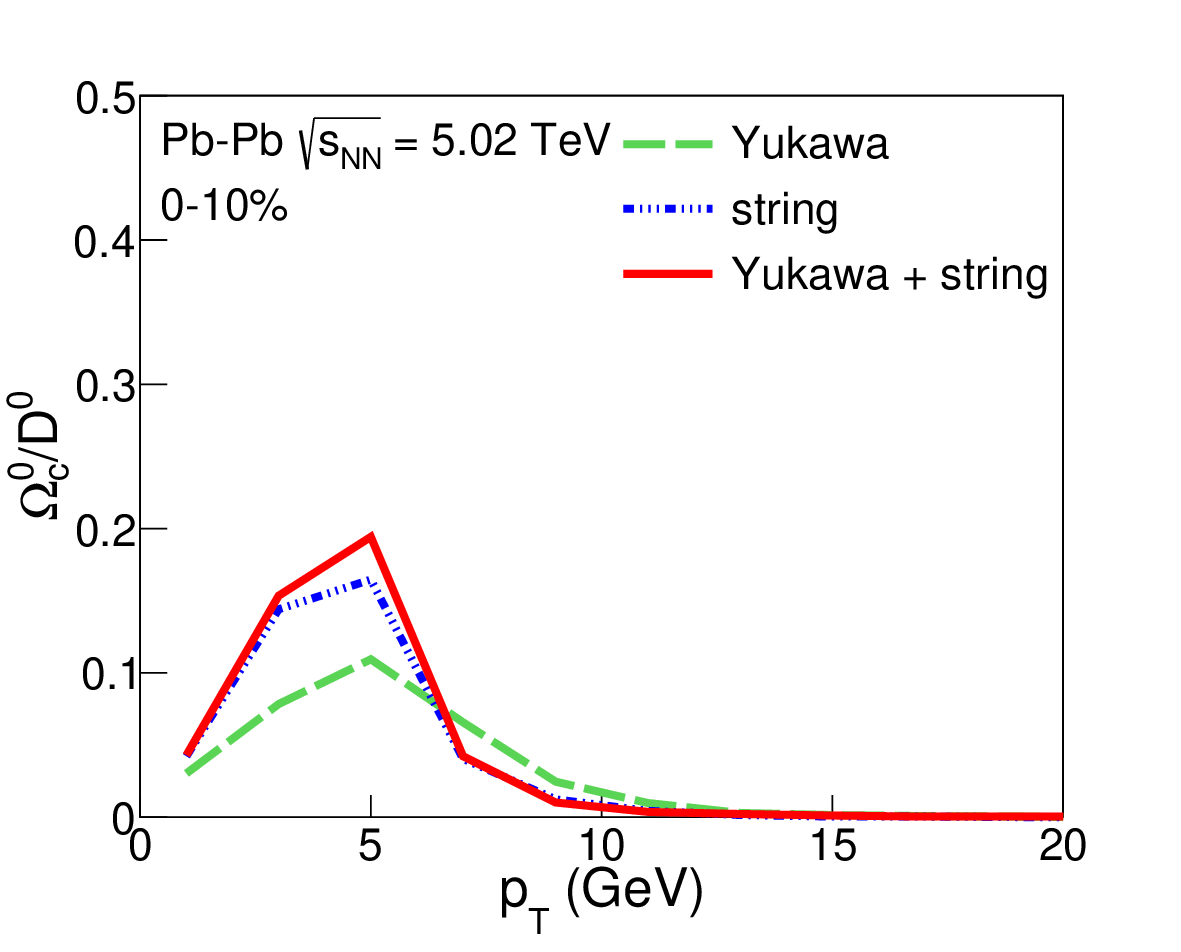}
    \caption{(Color online) The transverse momentum dependence of the $D_s^+/D_0$ (left panel), $\Lambda_c^+/D_0$ (middle panel), and $\Omega_c^0/D^0$ (right panel) ratios in 0-10\% Pb+Pb collisions at \(\sqrt{s_{\mathrm{NN}}}=5.02~\mathrm{TeV}\), compared between contributions from the Yukawa interaction, string interaction, and their combination. The ALICE data~\cite{ALICE:2022wpn} are shown for comparison.}
    \label{fig:Ds-Lamc-Ome-ratio}
\end{figure*}

\begin{figure*}[tbp]
   \centering
   \includegraphics[width=0.35\textwidth]{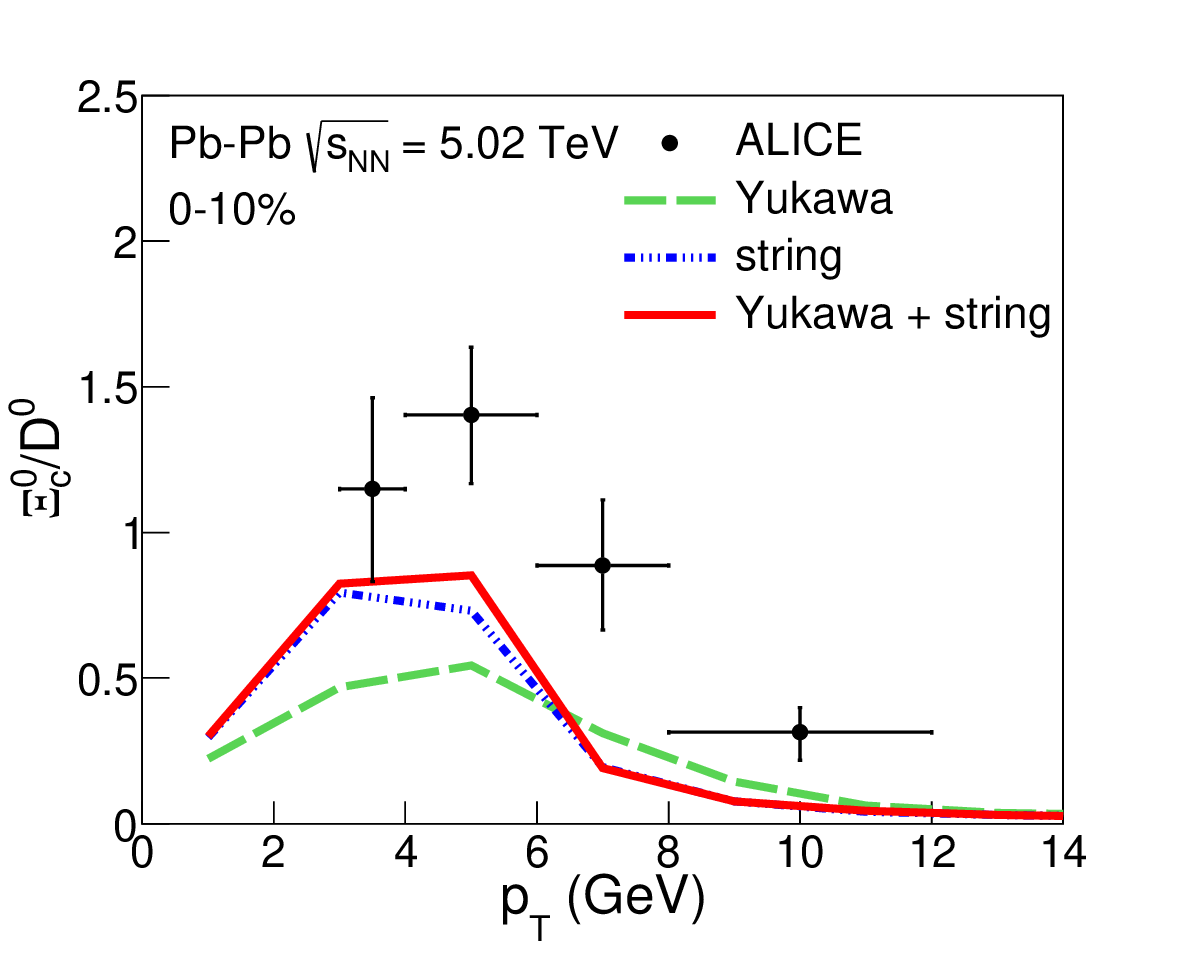}
   \hspace{-20pt}
   \includegraphics[width=0.35\textwidth]{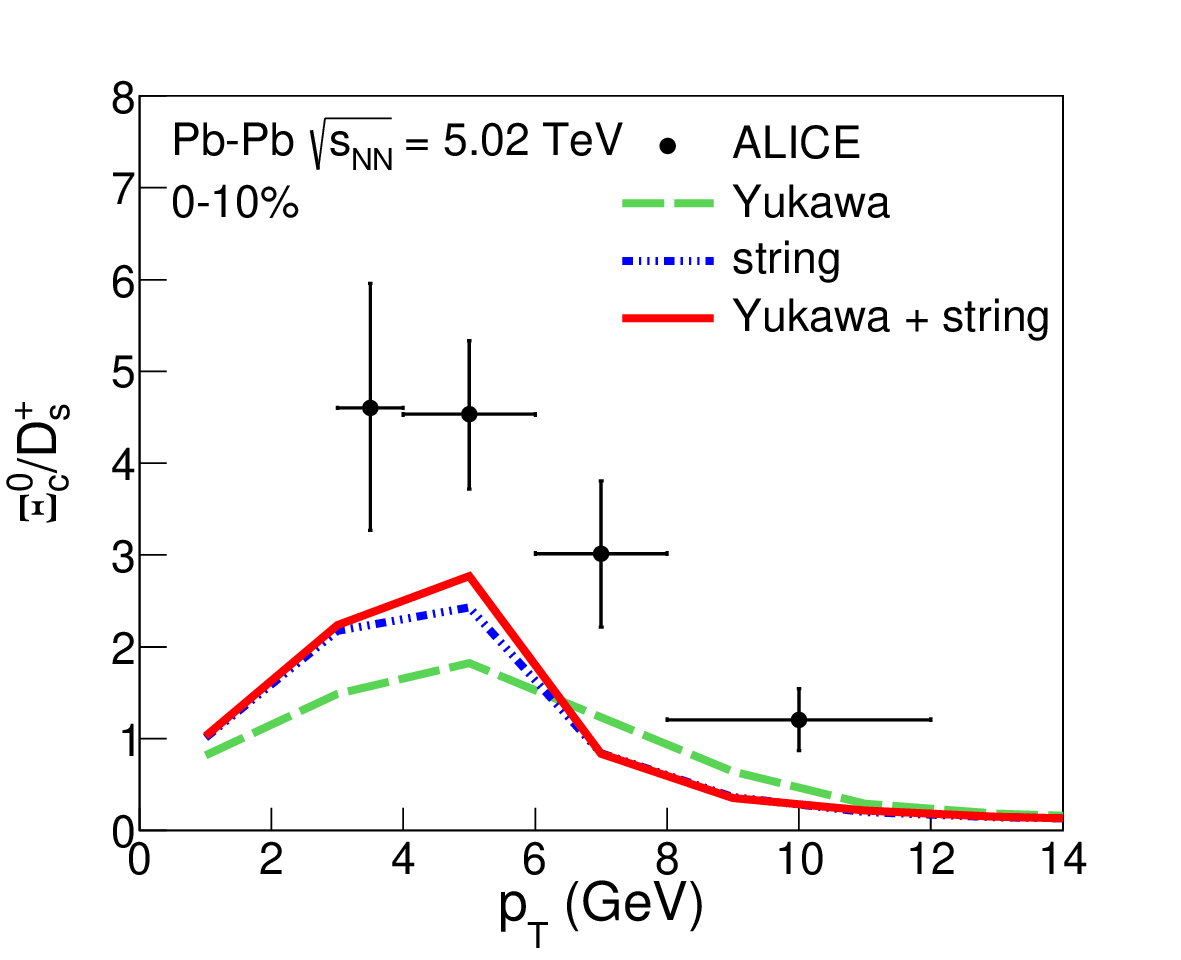}
   \hspace{-20pt}
   \includegraphics[width=0.35\textwidth]{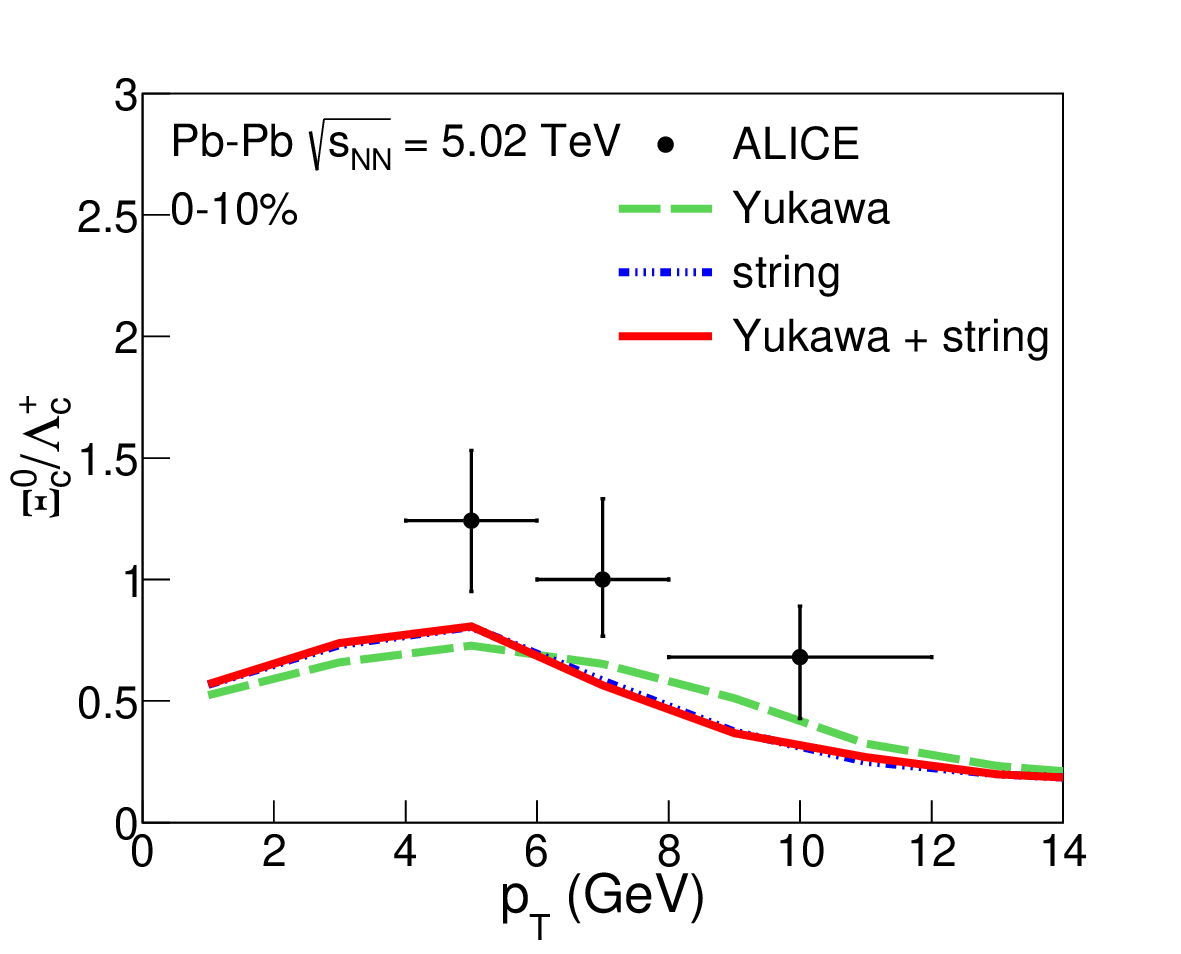}
   \caption{(Color online) The transverse momentum dependence of the $\Xi_c^0/D^0$ (left panel), $\Xi_c^0/D_s^+$ (middle panel), and $\Xi_c^0/\Lambda_c^+$ (right panel) ratios in 0-10\% Pb+Pb collisions at \(\sqrt{s_{\mathrm{NN}}}=5.02~\mathrm{TeV}\), compared between contributions from the Yukawa interaction, string interaction, and their combination. The ALICE data~\cite{ALICE:2026fzw} are shown for comparison.}
   \label{fig:Xic-ratio}
\end{figure*}

In Figs.~\ref{fig:Ds-Lamc-Ome-ratio} and~\ref{fig:Xic-ratio}, we present the ratios of spectra between different charm hadrons. Shown in Fig.~\ref{fig:Ds-Lamc-Ome-ratio} are the ratios of the \(D_s^+\), \(\Lambda_c^+\), and \(\Omega_c^0\) spectra to that of \(D^0\) in the 0-10\% Pb+Pb collisions at \(\sqrt{s_{\mathrm{NN}}}=5.02~\mathrm{TeV}\), compared between different interaction mechanisms and the ALICE data~\cite{ALICE:2022wpn}. All the ratios exhibit clear peak structures at intermediate $p_\mathrm{T}$ and then decrease towards fixed values at high $p_\mathrm{T}$. The strength of the peaks depends on the hadron species and the interaction mechanism: the peak value is largest when both the Yukawa and string interactions are included, and is smallest when only the Yukawa interaction is included. As previously discussed, this results from the enhanced contribution from coalescence in charm quark hadronization when charm quarks are closer to thermal equilibrium with the medium. The \(D_s^+/D^0\) ratio reflects the strangeness enhancement, the \(\Lambda_c^+/D^0\) ratio reflects the baryon enhancement, and the \(\Omega_c^0/D^0\) ratio reflects both. Since the coalescence probabilities to different charm hadrons rely on the relative velocity between a charm quark and a thermal quark, at low to intermediate $p_\mathrm{T}$ where coalescence dominates, the spectrum ratios between different hadrons strongly depend on the charm quark spectrum. On the other hand, the branching ratios to different hadrons in the Pythia fragmentation process have little dependence on the charm quark momentum. Therefore, at high $p_\mathrm{T}$ where fragmentation dominates, the spectrum ratios between different hadrons appear flat with respect to $p_\mathrm{T}$ and do not depend on the interaction mechanism between charm quarks and the QGP. We note that Pythia fragmentation does not provide reliable charm hadron chemistry in proton-proton collisions at low to intermediate $p_\mathrm{T}$~\cite{ALICE:2017thy,ALICE:2025wrq}. This has weak impact on our results in Pb+Pb collisions at low $p_\mathrm{T}$ where coalescence dominates, e.g., up to $p_\mathrm{T}\sim 5$~GeV for $D^0$ and $8$~GeV for $\Lambda_c$~\cite{Cao:2019iqs}, but would introduce a non-negligible uncertainty to our results at higher $p_\mathrm{T}$. 

The recent ALICE measurement on $\Xi_c^0$ in heavy-ion collisions~\cite{ALICE:2026fzw} provides an additional constraint on our understanding of the hadronization process of heavy quarks in the presence of a QGP medium. In Fig.~\ref{fig:Xic-ratio}, we present ratios of the \(\Xi_c^0\) spectrum to those of \(D^0\), \(D_s^+\), and \(\Lambda_c^+\) in 0-10\% Pb+Pb collisions at \(\sqrt{s_{\mathrm{NN}}}=5.02~\mathrm{TeV}\). Here, the \(\Xi_c^0/\Lambda_c^+\) ratio reflects the strangeness enhancement, the \(\Xi_c^0/D_s^+\) ratio reflects the baryon enhancement, and the \(\Xi_c^0/D^0\) ratio reflects both. The qualitative conclusions are the same as those for Fig.~\ref{fig:Ds-Lamc-Ome-ratio} above: the heavy-quark-medium coalescence produces the bump structures of these ratios at intermediate $p_\mathrm{T}$, signifying its enhanced production of strangeness and baryon number relative to the vacuum fragmentation. Hadron spectra formed by coalescence strongly depend on the medium-modified charm quark spectra. Compared to the Yukawa interaction, the string interaction causes stronger energy loss of charm quarks at low $p_\mathrm{T}$, drives charm quarks toward thermal equilibrium with the medium more rapidly, and therefore leads to stronger bump structures in these ratios via coalescence. To the contrary, these ratios are $p_\mathrm{T}$-independent at high $p_\mathrm{T}$ where fragmentation dominates. Although our calculation provides a better description of the experimental data than the other model predictions quoted in the ALICE publication~\cite{ALICE:2026fzw}, clear underestimation of these ratios still remains. This can result from the inaccurate branching ratios to different hadron species in the current Pythia fragmentation module as discussed earlier, and may also be improved by including decays from higher resonant states of charm hadrons beyond the current listings of the Particle Data Group~\cite{He:2019tik} in both coalescence and fragmentation. These will be explored in our future effort.


\section{Summary}
\label{sec:summary}

We have studied the nuclear modification of charm hadrons in heavy-ion collisions within a Langevin framework that incorporates both elastic and inelastic energy loss of heavy quarks inside the QGP. The Langevin model has been developed to accommodate a heavy quark transport coefficient that depends on both the heavy quark momentum and the medium temperature. The momentum space diffusion coefficient of heavy quarks is evaluated by a potential scattering model, in which both the short-range Yukawa (perturbative) and long-range string (non-perturbative) interactions between heavy quarks and the QGP are included. The string interaction dominates the heavy quark dynamics at low momentum and low temperature, while the Yukawa interaction dominates at high momentum and high temperature, as reflected by the momentum and temperature dependence of the spatial diffusion coefficient and drag coefficient of charm quarks. Inside a static medium, we have verified that a heavy quark system can approach the correct thermal equilibrium limit within this Langevin model. The string interaction generates a much higher thermalization speed than the Yukawa interaction does.

By coupling the improved Langevin model with an initial heavy quark spectrum from the FONLL calculation, a QGP medium simulated using the CLVisc hydrodynamic model, and a hadronization scheme that includes both coalescence and fragmentation mechanisms, we have investigated the nuclear modification effects on heavy flavor hadrons in realistic heavy-ion collisions. We have found that the Yukawa interaction dominates the heavy flavor observables at high \(p_\mathrm{T}\), while the string interaction dominates at low \(p_\mathrm{T}\). Their combination provides a reasonable description of the nuclear modification factor and elliptic flow coefficient of $D$ mesons at RHIC and the LHC. Since the string interaction increases rapidly as the medium temperature decreases towards the hadronization temperature \(T_\mathrm{c}\), including the string interaction allows heavy quarks to inherit stronger $v_2$ from the late stage of the QGP evolution when the anisotropic flow of the medium is fully developed. Therefore, compared to our earlier model setup with a constant spatial diffusion---$D_\mathrm{s}(2\pi T)$---of heavy quarks, the current setup provides a much stronger $v_2$ of $D$ mesons at low to intermediate $p_\mathrm{T}$. The temperature dependence embedded in the transport coefficient also allows us to describe the RHIC and LHC data using the same parameter setup. Effects of different interaction mechanisms are further reflected by the spectrum ratios between different charm hadrons, or the charm hadron chemistry. Compared to the Yukawa interaction, the string interaction drives charm quarks closer to thermal equilibrium with the medium and accumulates more charm quarks at low $p_\mathrm{T}$. This enhances the coalescence probability between charm quarks and thermal partons from the medium. Since the QGP medium is richer in quarks, in particular strange quarks, than in the particles produced by string breaking in vacuum, the enhanced coalescence process generates bump structures of the baryon-to-meson ratio and the strange-to-non-strange ratio at intermediate $p_\mathrm{T}$. These structures are consistently observed in our results on the \(D_s^+/D^0\), \(\Lambda_c^+/D^0\), \(\Omega_c^0/D^0\), \(\Xi_c^0/D^0\), \(\Xi_c^0/D_s^+\), and \(\Xi_c^0/\Lambda_c^+\) ratios. 

Results in this work are in quantitative agreement with those from the LBT model when the same Yukawa and string interactions are implemented~\cite{Xing:2021xwc}, validating a weak dependence of heavy quark phenomenology on the numerical implementation. Meanwhile, this Langevin model is much more efficient in computation than the LBT model, as long as one focuses on observables of single inclusive heavy flavor hadrons rather than their correlations with soft hadrons. This allows for a precise model calibration against the experimental data and a data-driven extraction of the heavy quark transport using Bayesian inference~\cite{Xu:2017obm,Xie:2022ght}, in which a rapid scan of the model calculation over a high-dimensional parameter space is required. This will be reported in an upcoming study. 

The numerical code developed for this work is publicly available in Ref.~\cite{githubLangevin}.


\section*{Acknowledgments}

This work is supported by the National Natural Science Foundation of China (NSFC) under Grant Nos.~12575146, 12505154, 12225503, 12321005 and~12175122, and by the Shandong Provincial Natural Science Foundation under Grant No.~ZR2025MS01.


\bibliographystyle{h-physrev5}
\bibliography{SCrefs}

\end{document}